\documentclass[
reprint,
superscriptaddress,
prl,
twocolumn,
amsmath,amssymb,
aps,
footinbib,
floatfix
]{revtex4-2}

\usepackage{graphicx}
\usepackage{xr-hyper}
\usepackage{hyperref}
\hypersetup{colorlinks=true, linkcolor=blue, citecolor=blue, urlcolor=blue}
\usepackage{xcolor}
\usepackage{lipsum}
\usepackage[caption=false]{subfig}
\usepackage{mathtools}
\usepackage[normalem]{ulem} % for \sout command

\usepackage[american]{babel}

\newcommand{\rbf}{\mathbf{r}}

\newcommand{\RI}{\sigma_I}

\newcommand\underrel[3][]{\mathrel{\mathop{#3}\limits_{%
			\ifx c#1\relax\mathclap{#2}\else#2\fi}}}

\renewcommand{\epsilon}{\varepsilon}

\newcommand{\gAA}{J_{AA}}
\newcommand{\gAB}{J_{AB}}

\begin{document}

\title{Flocking Beyond One Species: Novel Phase Coexistence in a Generalized Two-Species Vicsek Model}

\date{\today}

\begin{abstract}
\noindent A hallmark in natural systems, self-organization often stems from very simple interaction rules between individual agents. While single-species self-propelled particle systems are well understood, the behavior of binary mixtures with general alignment interactions remains largely unexplored with a few scattered results hinting at the existence of a rich emergent phase behavior. Here, we investigate systematically a generalization of the two-species Vicsek model with reciprocal intra- and interspecies (anti)alignment couplings, uncovering a rich phenomenology of emergent states. Notably, we show that rather than destroying polar order, antialigning interactions can promote phase separation and the emergence of global polar order. In doing so, we uncover a novel mechanism for microphase separation. We further find these coexistence patterns can be generalized to multispecies systems with cyclic alignment interactions.
\end{abstract}

\author{Eloise Lardet}
\author{Letian Chen}
\author{Thibault Bertrand}
\email[Electronic address: ]{t.bertrand@imperial.ac.uk}
\affiliation{Department of Mathematics, Imperial College London,
180 Queen’s Gate, London SW7 2AZ, United Kingdom}

\maketitle

%%%%%%%%%
% Introduction   %
%%%%%%%%%
For the past three decades, active matter has served as a playground for novel nonequilibrium emergent phenomena, where simple local interactions drive unexpected collective behaviors \cite{marchetti2013,ramaswamy2010}. It also offers a quantitative approach to complex biological processes. Biological systems—ranging from animal herds to bacterial suspensions and cytoskeletal filaments—remarkably self-organize into coherent structures despite nature's inherent stochasticity \cite{giardina2008,vicsek2012,peruani2012,cavagna2018,huber2018, peled2021,ariel2022}. A prototypical example of this is flocking, where local alignment interactions among self-propelled agents gives rise to macroscopic directed motion and true long-range order \cite{vicsek1995, toner1995, toner1998,tu1998,aditi2002,ramaswamy2003,toner2005,chate2008a,chate2008b,toner2012a,toner2012b, vicsek2012,cavagna2015,mahault2019, chate2020} at odds with the Mermin-Wagner-Hohenberg theorem \cite{mermin1966,hohenberg1967}.

Despite being under intense scrutiny since its inception three decades ago, Vicsek-like models still offer surprises \cite{toner2012a,solon2015,chepizhko2021,jentsch2024,yuan2024,zhao2025}, with even small modifications leading to novel mechanisms for long-range order \cite{chatterjee2023, lardet2024}. Shown to be more robust but also more fragile than originally expected, flocks are unstable to small perturbations like the presence of obstacles \cite{chepizhko2013,morin2017,codina2022}, spatial anisotropy \cite{solon2022} or even dissenters \cite{baglietto2013,yllanes2017} but equally arise without explicit alignment interactions \cite{romanczuk2009,deseigne2012,knezevic2022,casiulis2022,caprini2023,das2024,baconnier2025}. Beyond the usual polar alignment rules, recent studies have explored the emergence of collective motion with nematic alignment \cite{chate2006}, antipolar alignment \cite{escaff2024a,escaff2024b}, mixed alignment rules \cite{grossmann2015, denk2020} and in the presence of disorder \cite{peruani2018,toner2018a, toner2018b,duan2021,chen2022a,chen2022b,chen2022c,lardet2024}

Although foundational theoretical work in active matter has primarily focused on single-species systems, recent interest has turned to multispecies systems with more intricate interaction rules \cite{menzel2012,chatterjee2023,kursten2025,mangeat2025}. In particular, understanding the effect of nonreciprocal interactions in two-species systems has recently stolen the show \cite{fruchart2021,zhang2023a,kreienkamp2024a,kreienkamp2024b, martin2025, tang2025}. Namely, nonreciprocity in the alignment---in which species A tends to align with species B, while species B antialigns with species A---leads to novel emergent behaviors such as (anti)parallel flocking \cite{chatterjee2023,menzel2012,mangeat2025} or chirality \cite{fruchart2021,chen2024,kreienkamp2025}. Interestingly, with the exception of \citet{kursten2025}, the vast majority of existing studies make the simplifying assumption of aligning (i.e., ferromagnetic-like) intraspecies interactions. As a consequence, a comprehensive study of a multispecies Vicsek-like model with generic (anti)alignment interactions is crucially lacking from the literature.

In this Letter, we explore a generalized two-species Vicsek model where self-propelled particles follow XY-like alignment rules with independent intra- and interspecies couplings. Mapping the phase space of reciprocal (anti)alignment interactions, we uncover a rich spectrum of flocking, antiparallel flocking, and coexistence states. Notably, we identify a novel microphase-separated state featuring stable periodic traveling bands, emerging from antiferromagnetic intraspecies and ferromagnetic interspecies interactions. We present a heuristic stability argument for this phase and demonstrate its persistence in multispecies systems with cyclic alignment coupling matrices.

%%%%%%
% Model  %
%%%%%%
\textit{Model---}We study a continuous-time version of the Vicsek model (also called the flying XY model) composed of $N$ self-propelled point-particles belonging to two species $\{A,B\}$ with distinct intra- and interspecies local alignment rules. Particles are placed in a two-dimensional periodic box of size $L\times L$ with a density of $\rho=N/L^2$. Here, particles move with constant speed $v_0$ along the self-propulsion direction $\hat{\mathbf{p}}_i=(\cos\theta_i, \sin\theta_i)$, which experiences rotational diffusion and a torque stemming from alignment interactions. The position $\mathbf{r}_i$ and orientation $\theta_i$ of particle $i$ at time $t$ are then governed by the following overdamped Langevin equations:
\begin{subequations}
	\begin{align}
		\dot{\mathbf{r}}_i &= v_0 \hat{\mathbf{p}}_i, \label{eq:langevin_r} \\
		\dot{\theta}_i &= \frac{1}{n_i} \sum_{j\in\mathcal{N}_i} J_{s(i), s(j)} \sin(\theta_j-\theta_i) + \sqrt{2D_r} \ \xi_i, \label{eq:langevin_theta}
	\end{align} \label{eq:langevin}
\end{subequations}
where $\xi_i(t)$ is a zero-mean, unit-variance Gaussian white noise, and $\mathcal{N}_i=\{j : |\mathbf{r}_i-\mathbf{r}_j|\leq \sigma_I, i\neq j\}$ is the set of particles interacting with particle $i$, i.e., particles within the radius of interaction $\sigma_I$. The torque is normalized by the number of interaction neighbors $n_i=|\mathcal{N}_i|$, also known as a mean-sine model \footnote{This model is referred to as the additive sine model in the absence of a normalization of the alignment interactions by the number of neighbors \cite{zhao2021,chepizhko2021}.}. The coupling strength between two particles is $J_{s(i), s(j)}$, where $s(i)$ is the species of particle $i$. This coupling may be aligning ($J_{s(i), s(j)}>0$) or antialigning ($J_{s(i), s(j)}<0$). Populations $A$ and $B$ are taken to be in equal proportion such that there are $N/2$ particles of each species.
For the purposes of this Letter, we will assume that the intraspecies couplings are identical across both species $J_{\rm AA}=J_{\rm BB}$, and the interspecies couplings are equal $J_{\rm AB}=J_{\rm BA}$ leading to pairwise reciprocal interactions \footnote{Note that the interaction torques themselves may not be reciprocal; as particles may have different numbers of interaction neighbors $n_i$, the alignment interactions are generically nonreciprocal in this mean-sine alignment rules. Refer to \cite{chepizhko2021} for a discussion of this difference.}. We numerically solve Eq.\,(\ref{eq:langevin}) using an Euler-Maruyama method with time step $\Delta t=0.01$. We set $v_0=1$ and $\sigma_I=1$ unless stated otherwise, leaving the following free parameters: $\gAA, \gAB, D_r, \rho$, and $N$.

\begin{figure*}[!t]
\includegraphics[width=\linewidth]{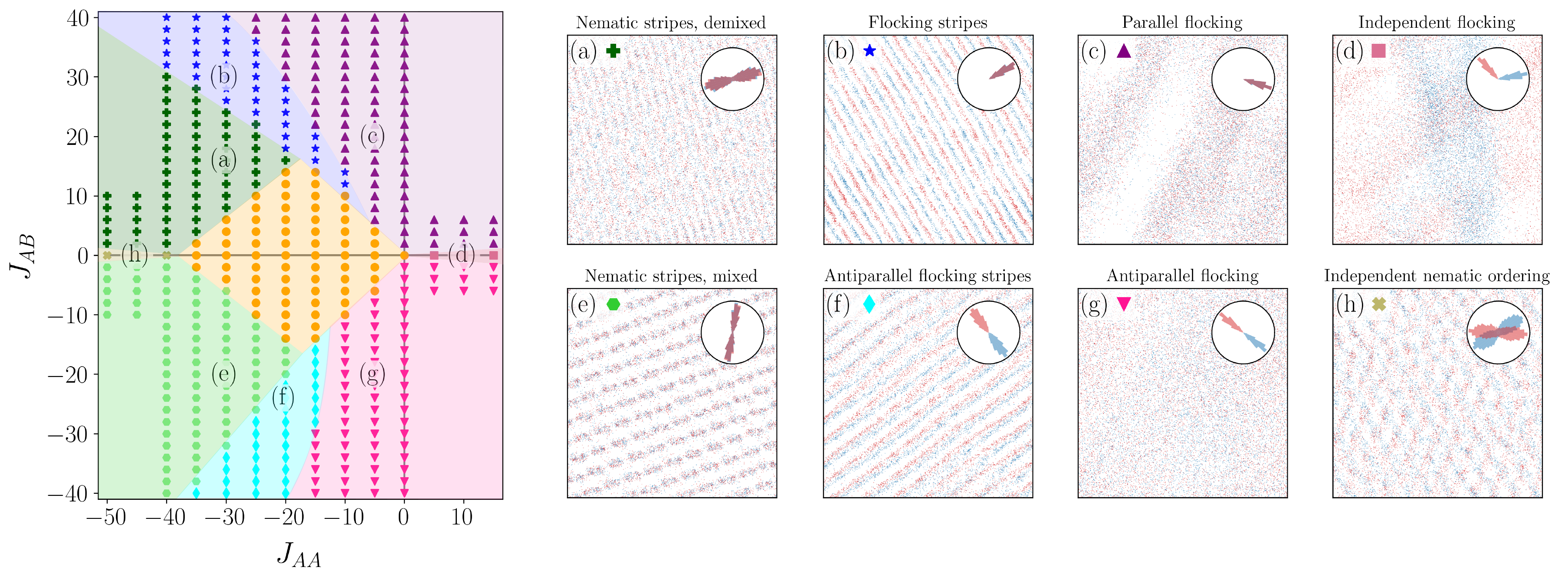}
\caption{Numerical phase diagram obtained from particle simulations. Here, we use the following parameters $N=2\times10^4$, $\rho=100$, and $D_r=0.2$. Phases were classified numerically, with data averaged over 10 independent realizations for each data point. Here, we list the phases observed and the parameters used to obtain the associated example snapshots:
(a) nematic stripes, demixed (dark-green plus), $\gAA=-30$, $\gAB=16$; (b) flocking stripes (dark-blue star), $\gAA=-30$, $\gAB=30$; (c) parallel flocking (purple upward triangle), $\gAA=-5$, $\gAB=20$; (d) independent flocking (pink square), $\gAA=10$, $\gAB=0$; (e) nematic stripes, mixed (light-green hexagon), $\gAA=-30$, $\gAB=-20$; (f) antiparallel flocking stripes (light-blue diamond), $\gAA=-20$, $\gAB=-24$; (g) antiparallel flocking (pink downward triangle), $\gAA=-5$, $\gAB=-20$; (h) independent nematic ordering (khaki cross), $\gAA=-45$, $\gAB=0$. Finally, we also observe a disordered hyperuniform phase (orange circles). Phase boundaries were added manually as a visual guide. See accompanying Videos S1--S9 and their descriptions in Supplemental Material \cite{Note3}.}
\label{fig:phase_diagram}
\end{figure*}

\begin{figure}
\includegraphics[width=\linewidth]{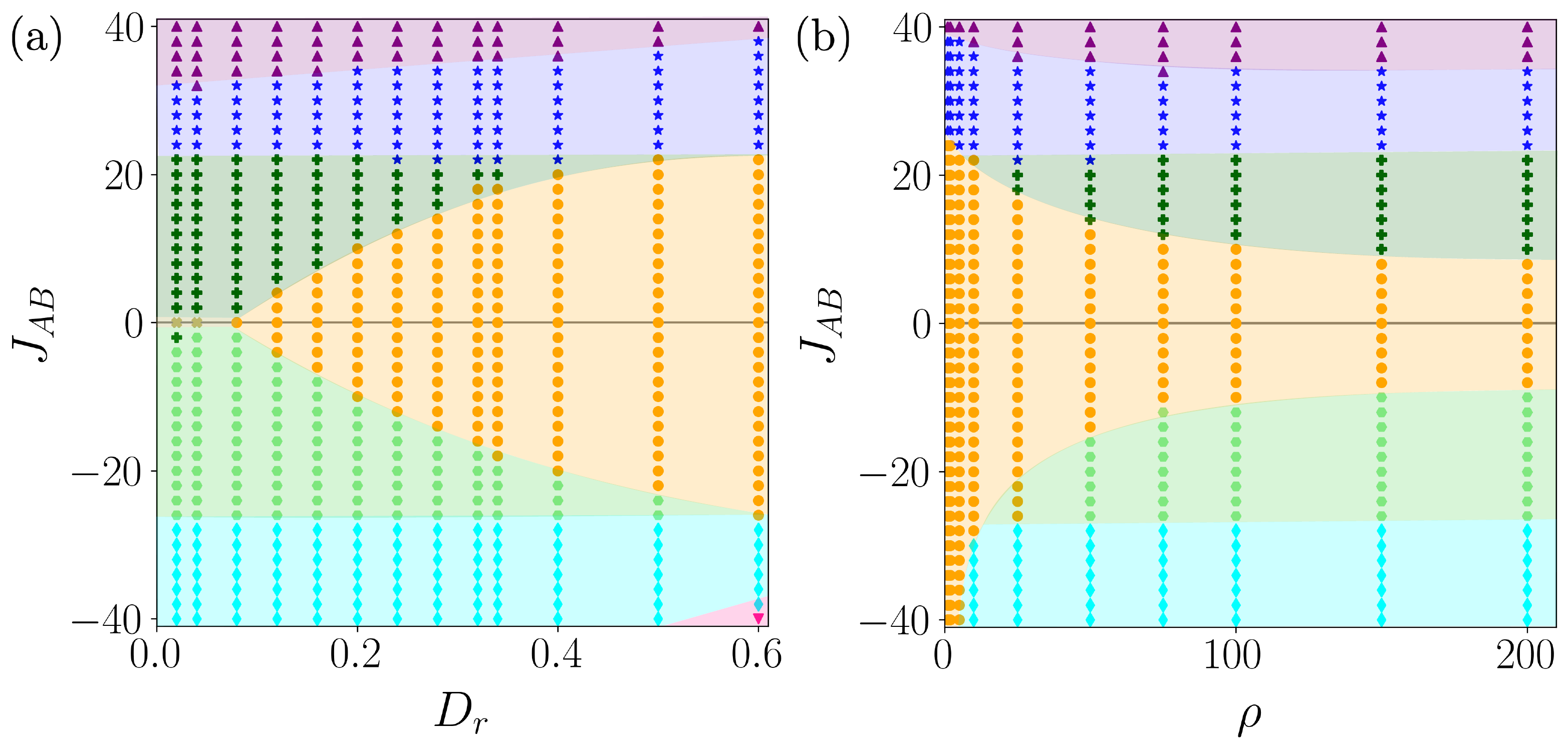}
\caption{Phase diagrams when varying noise strength and density. (a) Phase behavior in the $(\gAB, D_r)$ plane at fixed density $\rho=100$, intraspecies interaction strength $\gAA=-25$. (b) Phase behavior in the $(\gAB, \rho)$ plane at fixed rotational diffusion $D_r=0.2$ and intraspecies interaction strength $\gAA=-25$. Here again, the system size was taken to be $N=2\times 10^4$. Symbols used here are the same as those introduced in Fig.\,\ref{fig:phase_diagram}; phase boundaries were added manually as a visual guide.}
\label{fig:other_phase_diagrams}
\end{figure}

%%%%%%%%%%%%%%%%%%%%%%%%%
% Rich emergent behavior and phase diagram. %
%%%%%%%%%%%%%%%%%%%%%%%%%
\textit{Rich emergent behavior and phase diagram---}We explore the phase space spanned by $(\gAA,\gAB)$ at fixed density and noise strength (see Fig.\,\ref{fig:phase_diagram}). Here, we first work at a large density to ensure system-spanning structures are observed. Note that the influence of density $\rho$ and noise strength $D_r$ are discussed below (see Fig.~\ref{fig:other_phase_diagrams}). 

We observe a diverse range of emergent behaviors, which we classify according to a number of order parameters including polar order, nematic order, demixing, and spatial periodicity of observed structures. Further details of how these order parameters were calculated can be found in the End Matter and Supplemental Material \footnote{See Supplemental Material at http://link.aps.org/supplemental/10.1103/lt6f-yjpd for further analytical and computational details, including videos, which includes Refs.~\cite{chate2006, torquato2018}.}. First, we examine the phases observed for $\gAA>0$. For $|\gAB|\ll\gAA$, the usual polar ordered flocking phase is observed with each species flocking independently [Fig.~\ref{fig:phase_diagram}(d) and Video S4 in Supplemental Material \cite{Note3}]. As $\gAB$ increases, the interspecies alignment becomes strong enough to coordinate the flocking directions of the two species and we observe a globally polar ordered phase [Fig.~\ref{fig:phase_diagram}(c) and Video S3 in Supplemental Material \cite{Note3}]. Conversely, in systems with intraspecies alignment ($\gAA>0$) but interspecies antialignment ($\gAB<0$), the two populations display independently global polar order, but are flocking in antiparallel directions [Fig.~\ref{fig:phase_diagram}(g) and Video S7 in Supplemental Material \cite{Note3}], leading to an absence of polar order at the system level. This phase was observed originally in the two-species Vicsek model \cite{chatterjee2023}. Interestingly, we observe that the parallel and antiparallel flocking phases [see Figs.~\ref{fig:phase_diagram}(c) and \ref{fig:phase_diagram}(g)] persist even in the case of intraspecies antialigning interactions ($\gAA<0$) provided that the interspecies (anti)alignment $\gAB$ is large enough. In this regime, we expect the large-scale phase behavior to be controlled by the interspecies interactions alone.

%%%%%%%%%%%%%%%%%%%%%
%%%%%%%%%%%%%%%%%%%%%

Strikingly, when $\gAB\approx-\gAA$, a novel phase emerges; this phase is characterized by traveling bands of alternating species exhibiting global polar order, and we refer to it as the \textit{flocking stripes phase} [Fig.~\ref{fig:phase_diagram}(b) and Video S2 in Supplemental Material \cite{Note3}]. Interestingly, despite strong antialigning intraspecies coupling, particles self-organize to form a high-density system-spanning bands structure within their own species. We observe that these bands---which are reminiscent of the banding phase commonly observed in Vicsek models but stem here from a very different mechanism---are spatially periodic and bands of different species are nonoverlapping. The interspecies alignment, along with this interspecies demixing, facilitates flocking with strong global polar order. This novel phase behavior is analyzed in more detail below.

Upon reflection about the $\gAB$ axis (i.e., in the regime where $\gAB\lesssim \gAA$), a similar flocking striped phase is found; however, in this case, the antialigning nature of the interspecies couplings leads to systems of bands for each species flocking in opposite directions [Fig.~\ref{fig:phase_diagram}(f)]. Although reminiscent of the homogeneous density antiparallel flocking state [Fig.~\ref{fig:phase_diagram}(g)], the system is now clustered; it displays a nontrivial global nematic order and is spatially periodic (in the direction of nematic order), with species-species demixing when averaged over time. We therefore call this the \textit{antiparallel flocking stripes} phase (see Video S6 in Supplemental Material \cite{Note3}).

For stronger intraspecies antialignment ($\gAA \ll -1$), we encounter three further periodically striped phases. In the regions $\gAB>0$ and $\gAB<0$, phases are shown to display nematic order for both species in the same direction [Figs.~\ref{fig:phase_diagram}(a) and \ref{fig:phase_diagram}(e)]. In both cases, the structure exhibits spatial periodicity, with regions of high-density stripes or lanes. However, for $\gAB>0$, species remain demixed when averaged over time (see Video S1 in Supplemental Material \cite{Note3}), whereas the particles form local mixed clusters for $\gAB<0$ (see Video S5 in Supplemental Material \cite{Note3}). We therefore call these phases \textit{nematic stripes (demixed)} and \textit{nematic stripes (mixed)}, respectively. When $\gAB\approx 0$ but $\gAA\ll-1$, the species behave independently as expected; nematic order is still observed within each population due to the very strong antialigning couplings (see Video S8 in Supplemental Material \cite{Note3}). Similar nematic lanes and spatial structuring were also reported in previous models involving antialigning couplings \cite{grossmann2015,escaff2024a,escaff2024b}.

Finally, for moderate value of intraspecies antialignment and interspecies interactions (either aligning or antialigning), we observe a homogeneous \textit{disordered} phase. While no polar or nematic order is observed in the disordered phase, we have found strong spatial structuring; indeed, as shown in Supplemental Material \cite{Note3}, this is consistent with hyperuniform configurations. As this has been put into question by recent work on a related lattice model \cite{boltz2025}, confirming this result would require further analytical work, which is outside the scope of the current paper.

While the diverse phases in Fig.\,\ref{fig:phase_diagram} naturally invite questions about transition orders and tricritical points, characterizing these remains a formidable challenge that took decades to resolve even for the simpler single-species model. We have added numerical order parameter plots across different phase boundaries to the End Matter to provide quantitative insight; however, a full characterization of these phase transitions, including the investigation of possible tricritical points, remains a major task for future research.

%%%%%%%%%%%%%%%%%%%%%
% Robustness of the observed phases. %
%%%%%%%%%%%%%%%%%%%%%
\textit{Robustness of the observed phases---}As discussed above, we chose to study the phase diagram for high density in order to get system-spanning patterns. Here, we confirm that these emergent phases were observed over a wide range of noise strengths $D_r$ and densities $\rho$ [Fig.~\ref{fig:other_phase_diagrams}]. Notably, we observe that the disordered regime shrinks upon decreasing rotational diffusion $D_r$.  Further, the parallel and antiparallel flocking stripes regions (blue stars and pink diamonds, respectively) persist across the wide range of noise strengths tested here. In fact, the flocking stripes region shows a weak expansion at higher $\gAB$ values at larger noise strengths. In both cases, intraspecies nematic order is favored when $|\gAB|$ is decreased. We expect the disordered phase to take over at larger noises, eventually disrupting the polar and antiparallel polar stripes phases. 

\begin{figure}
	\includegraphics[width=\linewidth]{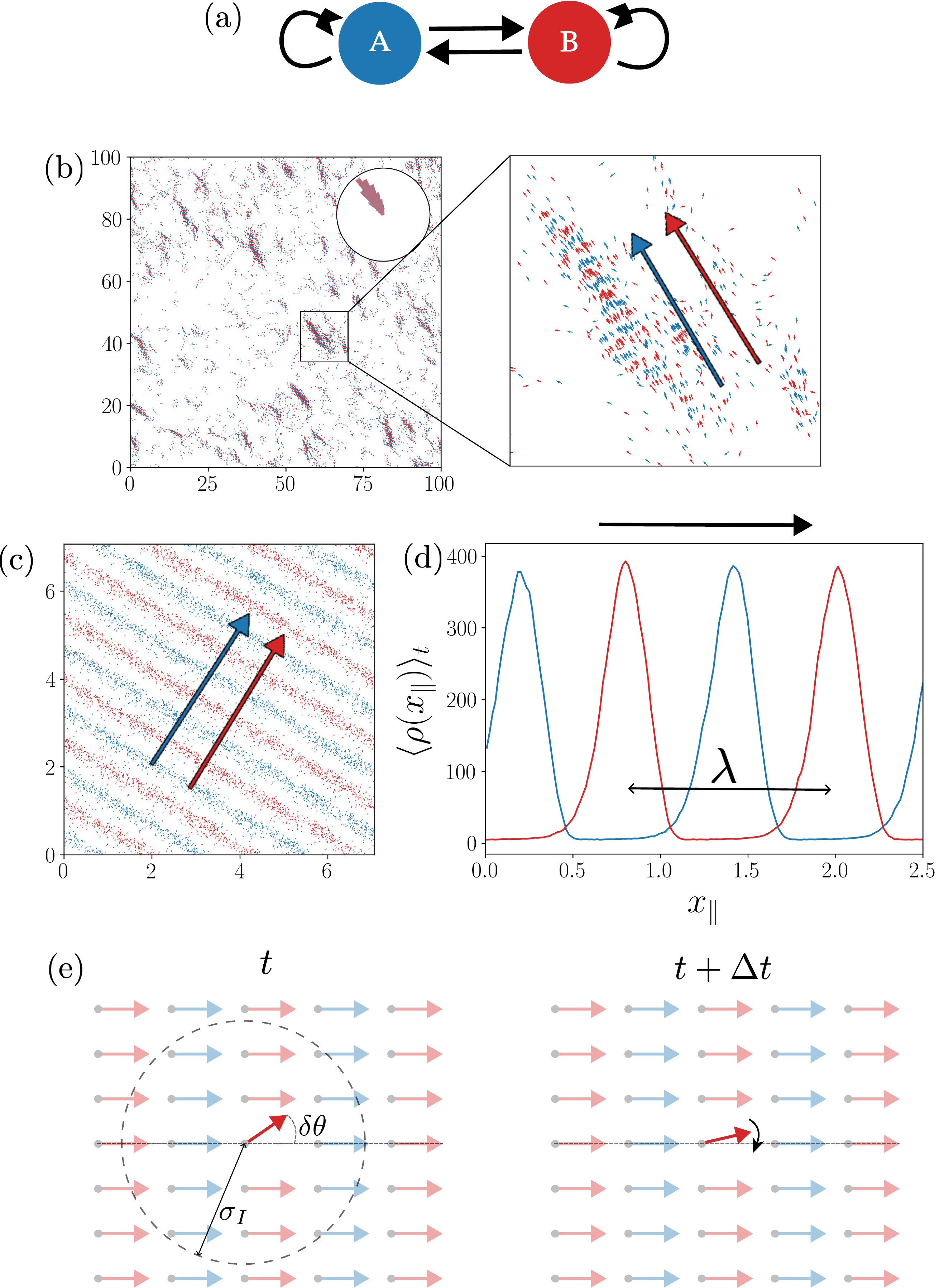}
	\caption{(a) Sketch of interactions between species in the flocking stripes phase. The interspecies couplings are aligning, whereas the intraspecies coupling is antialigning.
	(b) Snapshot of a simulation at low density for $\gAA=-\gAB=-20$ (parallel flocking with phase separation) with enlarged snapshot showing the microphase separation. The inset shows a polar histogram of the particle angles. These simulations were performed with $N=10^4$, $\rho=1$, and $D_r=0.02$. (c) Snapshot of a high-density simulation in the traveling band phase ($N=10^4$, $\rho=200$, $D_r=0.2$, $\gAA=-\gAB=-20$) and (d) its time-averaged density profile, projected along the direction of travel $x_\parallel$.
	(e) Simplified schematic to explain the stability mechanism.}
	\label{fig:stripes_snaps}
  \end{figure}

Examining the effect of density, the disordered region grows as $\rho$ decreases. This is likely due to the fact that the relative sparseness of these systems allows the particles to alleviate their competing alignment frustration through spatial distancing, rather than being forced to alleviate this through collective directed motion.
Just as in the case of varying $D_r$, the parallel and antiparallel flocking stripes phases persist for a remarkably wide range of densities.
For lower densities, the systems displaying flocking stripes tend to form isolated clusters of striped patterns displaying polar order [see Fig.~\ref{fig:stripes_snaps}(b)].
Predicting the phase boundaries as well as the nature of all these transitions analytically remains a formidable challenge and is the focus of future studies. 

%%%%%%%%%%%%%%%%%%%%%
%%%%%%%%%%%%%%%%%%%%%

Further, we tested the robustness of all observed phases to finite size effects by simulating a larger range of system sizes (see Supplemental Material \cite{Note3}). We found the phases to persist up to the largest system sizes simulated ($N=2\times10^5$). The flocking stripes phase is also robust to nonsymmetric parameter choices, such as unequal particle numbers or $\gAA\neq\gAB$ (see Supplemental Material \cite{Note3}).
Finally, we confirmed that our findings were robust to changes in the exact microscopic interaction rules; in particular, upon rescaling of the interaction couplings, we find that an additive sine model displays qualitatively the same phase diagram and confirmed that the flocking stripes patterns are also found in models with soft particle repulsion (see Supplemental Material \cite{Note3}). We note that interaction kernels that decay too fast tend to hinder flocking stripes (see Supplemental Material \cite{Note3}).

%%%%%%%%%%%%%%%%%%%%%
% Robustness of the observed phases. %
%%%%%%%%%%%%%%%%%%%%%
\textit{Characterization of the flocking stripes phase---}Finally, we focus for a moment on trying to characterize further the emergence and the stability of the flocking stripes phase, perhaps one of the most counterintuitive emergent behaviors presented here. As shown in Figs.\,\ref{fig:phase_diagram} and \ref{fig:other_phase_diagrams}, the flocking stripes phase appears over a wide range of parameters $\gAA, \gAB$. The two key ingredients giving rise to this phase are as follows: (1) one needs intraspecies antialignment ($\gAA<0$) but interspecies alignment ($\gAB >0$); (2) the intraspecies and interspecies interaction strengths should be similar in absolute value, $|\gAA| \approx |\gAB|$. For simplicity, we here focus on the case in which the couplings are equal in magnitude and of opposite signs. While particles align with the opposite species, they antialign within their own [Fig.\ref{fig:stripes_snaps}(a)]; it thus seems quite counterintuitive at first that particles would cluster and form dense bands within their own species. Although these bands resemble those in the coexistence regime of the two-species Vicsek model \cite{chatterjee2023}, we argue that this novel phase separation stems from a distinct physical mechanism.

At low densities, particles self-organize into finite polar clusters exhibiting stripes of alternating species as seen in Fig.~\ref{fig:stripes_snaps}(b) (see also Video S10 in Supplemental Material \cite{Note3}). Interestingly, this low-density clustered phase displays large global polar order. At high densities, the clusters become system-spanning traveling waves within the periodic box [Fig.~\ref{fig:stripes_snaps}(c)]. To characterize the stripes further, we project the particles onto the mean direction of travel and measure the density profile of the bands, averaged over time $\langle \rho(x_\parallel)\rangle_t$ [see Fig.~\ref{fig:stripes_snaps}(d)]. The stripes profiles are extremely regular, periodic in the direction of travel, and of equal amplitude forming an alternating smectic A pattern after a short transient. In the Vicsek model, traveling bands are found to eventually organize into a periodic pattern but only after very long transients \cite{chate2020}. Furthermore, in our case the local density between bands drops to almost zero, in contrast with the low-density gas state observed in the Vicsek model bands \cite{solon2015,solon2022}.

Measuring the wavelength between density peaks of the same species for various different interaction radii reveals that the wavelength is linearly dependent on the radius of interaction, with $\lambda\approx 1.23 \RI$, independently of the system density (see Supplemental Material \cite{Note3}). Particles thus form evenly spaced bands of alternating species, with the peaks of each species just over $\RI$ apart. The same wavelength also appears in the other patterns, and even in the disordered hyperuniform state (see Supplemental Material \cite{Note3}). Notably, we also found that the correlations in the velocity perpendicular to the direction of motion $\mathbf{v}^{(i)}_{\perp}$ decay extremely fast with time, pointing at the stability of this polar ordered phase.  

%%%%%%%%%%%%%%%%%%%%%
%%%%%%%%%%%%%%%%%%%%%

How are the flocking stripes so stable despite such strong antialignment between agents of the same species? To understand this, we devise the following mean-field argument: first assume without loss of generality that vertical stripes of alternating species have emerged, all with their polarization pointing in the $x$ direction (i.e., $\theta = 0$) as in Fig.~\ref{fig:stripes_snaps}(e) \footnote{Note that we only sketched a single column of particles for illustration purposes, when in reality there are potentially hundreds of particles clustered in each stripe.}. As already argued, the distance from the center of one stripe to another one of the same species is approximately $1.2\RI$; thus, the red particle highlighted in Fig.~\ref{fig:stripes_snaps}(e) only interacts with the particles in the same stripe (i.e., same species particles) and the two nearest neighbor stripes (i.e., opposite species particles) found directly in front and behind.

At time $t$, let us assume that red particle $i$ experiences a fluctuation in its orientation $\delta \theta>0$, due, for instance, to rotational diffusion. Because of the metric nature of the interactions, particle $i$ interacts with more particles of the other species $n_i^{\textrm{blue}}$ than with particles of its own species $n_i^{\textrm{red}}$. Denoting $\kappa=n_i^{\textrm{blue}}-n_i^{\textrm{red}} >0$ and $\gAB=-\gAA\equiv J$, then the torque on particle $i$ at time $t+\Delta t$ is given by $\mathcal{T}_i = -\frac{J \kappa}{n_i}\sin{\delta \theta}$, where $n_i = n_i^{\textrm{blue}}+n_i^{\textrm{red}}$. We conclude that $\delta \theta$ and $\mathcal{T}_i$ have opposite signs, which means that the torque is in the opposite direction to the direction of the original fluctuation and the particle's polarization is pulled back toward the mean direction of travel. The remaining particles experience the torque $\mathcal{T}_j=\pm \frac{J}{n_j}\sin{\delta \theta}$ (the sign depends on if it is a red or blue particle), which is much smaller in magnitude than $\mathcal{T}_i$. The timescale for particle $i$ to correct its orientation is therefore much faster than the change in orientation experienced by the neighbors. We conclude that the system is robust to orientational fluctuations and is self-stabilizing. 

\begin{figure}
	\includegraphics[width=\linewidth]{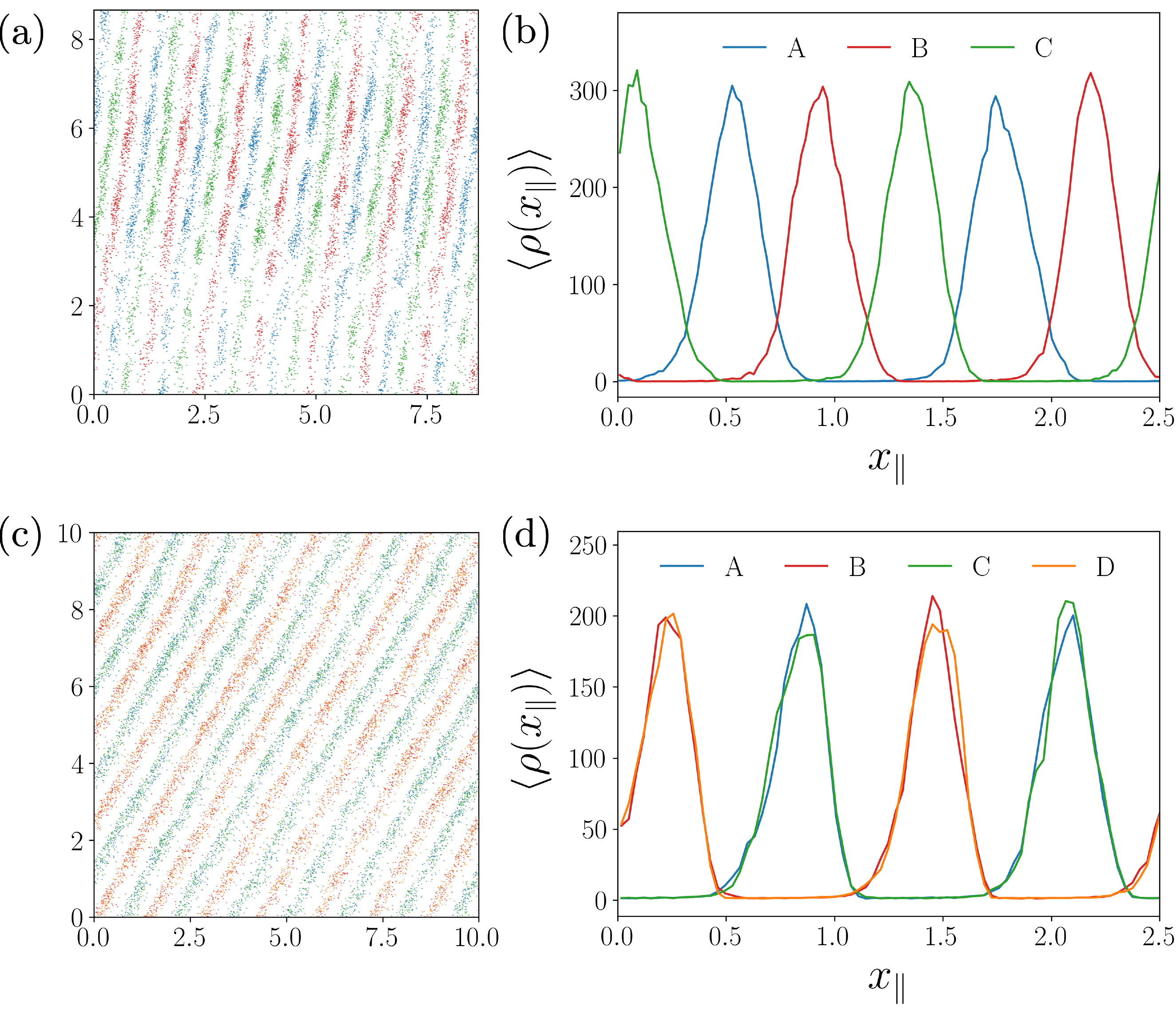}
	\caption{Multispecies ($m>2$) flocking. Snapshots of systems with alignment interactions governed by Eq.\,(\ref{eq:cyclic_multispecies}) with (a) $m=3$ and (c) $m=4$ species. (b)--(d) Time-averaged density profiles for the systems in (a)--(c), respectively. Simulation parameters were: $N=m\times 5\times 10^4$, $\rho=200$, $D_r=0.2$, $J=10(m+1)$.}
	\label{fig:multispecies}
  \end{figure}

\textit{Multiple species---}We investigate whether alternating flocking stripes can form for any number of species. Extending our model to $m$ species, we find that the flocking stripes phase persists under the following coupling rules:
\begin{equation}
	J_{ab} = \begin{cases}
		-J, & a=b, \\
		J, & b =(a+1) \mod m, \\
		0, & \textrm{otherwise},
	\end{cases}
	\label{eq:cyclic_multispecies}
\end{equation}
with $J>0$ and species numbered $a,b\in\{0,1,\dots,m-1\}$. As shown in Fig.~\ref{fig:multispecies}, for $m>2$, only odd $m$ produce $m$ distinct chasing stripes, while even $m$ leads to overlap with every second species, effectively reducing the system to the $m=2$ case. A similar parity dependence was previously observed in attraction-repulsion cyclic interactions \cite{ouazan-reboul2023}. The behavior of multispecies ($m>2$) Vicsek models under general interactions presents intriguing open questions, which we will address in future studies.

%%%%%%%%%%%%%%%%%%%%%
%%%%%%%%%%%%%%%%%%%%%

\textit{Conclusion and outlook---}Despite being under intense scrutiny over the past three decades, Vicsek-like models still offer surprises. Recent studies have shown that multispecies systems of self-propelled particles can generate intricate patterns \cite{ouazan-reboul2023,dinelli2023}, including moving stripe and cluster formations under both reciprocal \cite{chatterjee2023,menzel2012,mangeat2025} and nonreciprocal \cite{fruchart2021,zhang2023a,kreienkamp2024a,kreienkamp2024b,martin2025,tang2025,saha2020,you2020,alston2023} interactions. In this Letter, we have provided a comprehensive study of a generalized two-species Vicsek model with homogeneous and reciprocal couplings. Here, purely reciprocal alignment has led to a novel mechanism for interspecies phase separation and the emergence of collective motion. Distinct (anti)parallel flocking and nematic flocking phases were identified along the way. In particular, we highlighted a novel flocking stripes phase as a striking example of how alignment across species, yet antialignment within a species, can lead to a counterintuitive stable traveling bands state.

A theoretical approach to this problem is addressed in a companion paper \cite{lardet2025}, in which we derive a kinetic framework for the model and perform linear stability analysis. The theoretical results agree well with our particle-based simulation results. Strikingly, they also predict a finite wavelength instability with corresponding wavelength $\lambda=1.23$ for the novel coexistence phases. The analytical results also explain the differences in parity for $m>2$ species flocking stripes and their stripe ordering.
The aforementioned extension of this problem to general interaction networks remains an open challenge that we will address in future work. We expect this to be fundamental to our understanding of the connection between the structure of the interaction couplings matrix and the resulting emergent behavior, e.g., in drawing parallels with other cyclic interaction models \cite{szolnoki2020, ouazan-reboul2023, ouazan-reboul2023a,ouazan-reboul2023b}. By uncovering how global polar order emerges directly from species-level antialignment, we provide a new theoretical framework to interpret observed phase separation in coexisting biological populations. These robust ``flocking stripe" and nematic phases offer an interesting set of rules that serve as a blueprint for controlling emergent behavior in synthetic systems, including programmable robotic swarms or active colloids with specifically designed interspecies interactions.

\textit{Note added---}A preprint independently reporting the flocking stripes phase within a similar model \cite{oki2025} was released during the finalization of this manuscript.

\textit{Acknowledgments---}The authors would like to thank Emir Sezik and Jacob Knight for useful discussions. E.~L. was funded by a President's PhD Scholarship at Imperial College London. The authors acknowledge computing resources provided by the Imperial College Research Computing Service.

\textit{Data availability---}The data that support the findings of this article are openly available \cite{lardet2026}.

%%%%%%%%%%%%%%%%%%%%%
% \bibliography{refs_two_pop.bib}


\appendix*
\clearpage
\section{End Matter}
Here we provide a brief overview of the quantities used to classify the phases in Figs.~\ref{fig:phase_diagram} and \ref{fig:other_phase_diagrams}. Further details on the calculation of the order parameters and classification of phases can be found in Supplemental Material \cite{Note3}.

For a set $\Omega$ of particles, the global \emph{polar order} over the set $\Omega$ is given by
\begin{equation}
	\Psi_{\Omega} = \left \langle \frac{1}{|\Omega|} \left|\sum_{i \in \Omega} \hat{\mathbf{p}}_i(t) \right| \right \rangle,
\end{equation}
where $|\Omega|$ is the number of particles in the set $\Omega$ and the angular brackets denote an average over time, once in steady state. We measure the polar order within each species $\Psi_A$, $\Psi_B$, as well as the overall polar order $\Psi$. We also measure the difference in polar direction between the two species $\delta\theta_\Psi$.

Similarly, we can measure the \emph{nematic order} parameters $S_A$, $S_B$, and $S$ using
\begin{equation}
	S_{\Omega} = \left \langle \frac{1}{|\Omega|}\left|\sum_{i \in \Omega} \hat{\mathbf{q}}_i(t) \right| \right \rangle,
\end{equation}
where $\hat{\mathbf{q}}_i(t)=(\cos2\theta_i(t), \sin2\theta_i(t))$ is the unit nematic orientation vector and $|\Omega|$ is the number of particles in the set $\Omega$. We also measure the difference in nematic direction between the two species $\delta\theta_S$.

The degree of local species phase separation is measured via the following \emph{demixing order parameter}:
\begin{equation}
\bar{d} = \left \langle \frac{1}{N} \sum_{i=1}^N \frac{1}{|\mathcal{D}_i|} \sum_{j\in\mathcal{D}_i} \delta_{s(i),s(j)}-\frac{1}{2}\right \rangle,
\end{equation} 
where $\mathcal{D}_i=\{j:|\rbf_i-\rbf_j|\leq r_d\}$ and $\delta_{s(i)s(j)}$ is a Kronecker delta.
This quantity measures the average ratio of same species particles to total number of neighbors within a certain distance $r_d$ of each particle. Note that we subtract a $1/2$ such that $\bar{d}>0$ describes a demixed binary system.
We are concerned with demixing that occurs on a length scale smaller than the interaction radius; thus, we choose $r_d=\RI/2$ so that $\mathcal{D}_i$ is almost only comprised of particles of the same species in the striped phases. The value of $r_d$ could be tuned further, but we found this to be sufficiently sensitive to changes in species demixing.

Finally, for each system, we quantify the periodicity of spatial structures by checking for the appearance of a large peak around $k\approx1/\lambda\approx 0.8$ in the power spectrum $|\delta\hat{\rho}(k)|^2$ of the density fluctuations parallel to the direction of order. This is a particularly helpful measure when the demixing order parameter is less sensitive to phase changes, for instance, between antiparallel flocking (pink downward triangle) and antiparallel flocking stripes (light-blue diamond).

\begin{figure}
\includegraphics[width=\linewidth]{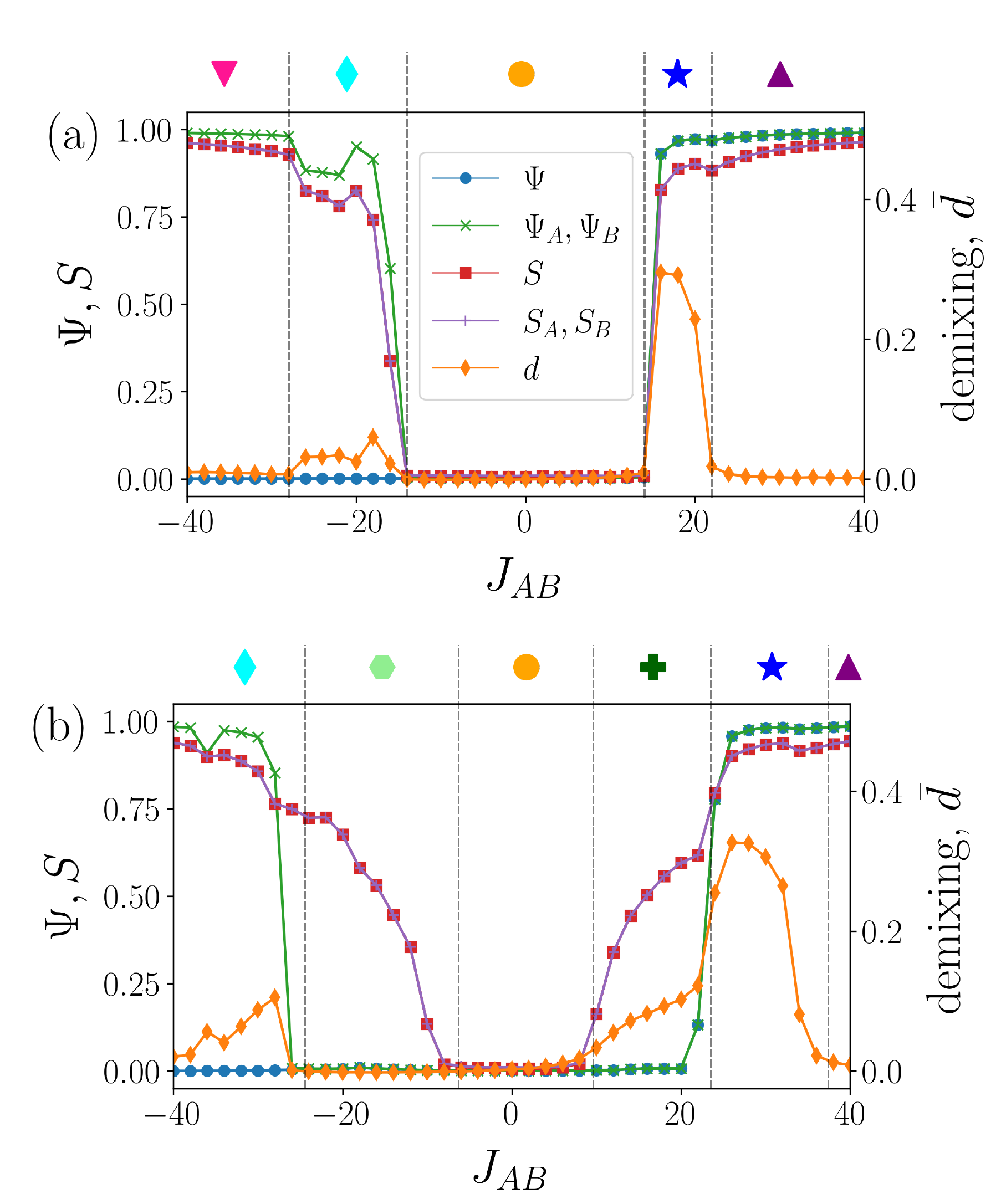}
\caption{Numerical order parameters (polar order, nematic order, and demixing) against $\gAB$ for fixed (a) $\gAA=-15$ and (b) $\gAA=-25$. Vertical gray dashed lines show the phase boundaries between different phases, whose symbols plotted above correspond to those in Fig.~\ref{fig:phase_diagram}.}
\label{fig:order_params}
\end{figure}

The numerical order parameters are plotted in Fig.\,\ref{fig:order_params} along the vertical slices at $\gAA=-15$ and $\gAA=-25$ from Fig.\,\ref{fig:phase_diagram}. A more refined characterization of the phase transitions is left as future work.

%%%%%%%%%%%%%%%%%%%%%
%%%%%%%%%%%%%%%%%%%%%

\end{document}

\title{Supplementary Material for ``Flocking Beyond One Species: Novel Phase Coexistence in a Generalized Two-Species Vicsek Model''}
\date{\today}

\author{Eloise Lardet}
\author{Letian Chen}
\author{Thibault Bertrand}
\email[Electronic address: ]{t.bertrand@imperial.ac.uk}
\affiliation{Department of Mathematics, Imperial College London,
180 Queen’s Gate, London SW7 2AZ, United Kingdom}

\maketitle

\hrule height 0.5mm

{\hypersetup{linkcolor=black}
\tableofcontents}

\vspace{2em}
\hrule height 0.5mm

\section{Supplementary movies}
We provide supplementary movies to accompany the snapshots in Fig.~1 of the main text, along with an additional movie in the disordered state. The parameters used are given as follows:

\begin{itemize}
	\item Movie S1 (nematic stripes, demixed): $N=2\times10^4$, $\rho=100$, $D_r=0.2$, $\gAA=-30$, and $\gAB=16$.
	\item Movie S2 (flocking stripes): $N=2\times10^4$, $\rho=100$, $D_r=0.2$, $\gAA=-30$, and $\gAB=30$.
	\item Movie S3 (parallel flocking): $N=2\times10^4$, $\rho=100$, $D_r=0.2$, $\gAA=-5$, and $\gAB=20$.
	\item Movie S4 (independent flocking): $N=2\times10^4$, $\rho=100$, $D_r=0.2$, $\gAA=10$, and $\gAB=0$.
	\item Movie S5 (nematic stripes, mixed): $N=2\times10^4$, $\rho=100$, $D_r=0.2$, $\gAA=-30$, and $\gAB=-20$.
	\item Movie S6 (antiparallel flocking stripes): $N=2\times10^4$, $\rho=100$, $D_r=0.2$, $\gAA=-20$, and $\gAB=-24$.
	\item Movie S7 (antiparallel flocking): $N=2\times10^4$, $\rho=100$, $D_r=0.2$, $\gAA=-5$, and $\gAB=-20$.
	\item Movie S8 (independent nematic ordering): $N=2\times10^4$, $\rho=100$, $D_r=0.2$, $\gAA=-45$, and $\gAB=0$.
	\item Movie S9 (disordered): $N=2\times10^4$, $\rho=100$, $D_r=0.2$, $\gAA=-20$, and $\gAB=4$.
	\item Movie S10 (low density flocking stripes): $N=10^4$, $\rho=1$, $D_r=0.02$, $\gAA=-20$, and $\gAB=20$.
\end{itemize}

\section{Classification}

We explored the phase space and classify observed phases based on the following order parameters:

\begin{itemize}
	\item Polar order (globally $\Psi$ and within each population $\Psi_A, \Psi_B$);
	\item Nematic order (globally $S$ and within each population $S_A$, $S_B$);
	\item Difference in direction of polar order $\delta \theta_{\Psi}$ and nematic order $\delta \theta_{S}$ between the two species; 
	\item Demixing order parameter $d$;
	\item Periodicity in direction of polar (or nematic) order.
\end{itemize}
Below, we provide definitions for each of these order parameters. Table~\ref{tab:classification} summarizes how each phase is uniquely determined.

\subsection{Polar order parameter}
For a set $\Omega$ of particles, the polar order of the system is given by
\begin{equation}
	\psi_{\Omega}(t) = \frac{1}{|\Omega|} \left|\sum_{i \in \Omega} \hat{\mathbf{p}}_i(t) \right|,
\end{equation}
where $|\Omega|$ is the number of particles in the set $\Omega$. This quantity is then averaged over time, once in steady state: $\Psi_{\Omega} = \langle \psi_{\Omega}(t)\rangle_t$. We measure the polar order over all particles, $\Psi$, and within each species $\Psi_A$ and $\Psi_B$.

For the special case when species A and species B flock independently (e.g. $J_{AA}>0, J_{AB}\approx0$), although there is no globally preferred direction, we may observe $\Psi>0$, above the flocking threshold, due to the strong flocking of each species. However, in general the direction of species A and B will not be the same, hence we measure the difference in their preferred direction as follows: define $\alpha = \theta_A - \theta_B ~(\mathrm{mod}~2\pi)$ where $\theta_\Omega=\langle \theta_\Omega(t)\rangle_t$ is the time-averaged mean direction of particles in the set $\Omega$. All angular averages were computed by applying $\mathrm{atan2}$ to the vectorial averages of unit vectors $\pbf_i=(\sin\theta_i, \cos\theta_i)$ to preserve periodicity. Then the difference in polar direction is $\delta \theta_{\Psi} = \min\{\alpha, 2\pi-\alpha\}$ to ensure that $\delta\theta_\Psi \in [0,\pi)$.

\subsection{Nematic order parameter}
Similarly, for a set of particles $\Omega$, the nematic order over $\Omega$ is given by 
\begin{equation}
	S_{\Omega}(t) = 2 \,\sqrt{\left(\frac{1}{|\Omega|}\sum_{i \in \Omega} \cos^2{\theta_i(t)} - \frac{1}{2}\right)^2 + \left(\frac{1}{|\Omega|}\sum_{i \in \Omega} \cos\theta_i(t)\sin\theta_i(t) \right)^2}
\end{equation}
where $|\Omega|$ is the number of particles in the set $\Omega$ \cite{chate2006}.
This quantity is then averaged over time, once it reaches a statistically stationary state: $S_{\Omega} \equiv \langle S_{\Omega}(t)\rangle_t$.
We measure the polar order over all particles, $S$, and within each species giving $S_A$ and $S_B$.

For the special case when species A and species B show nematic order independently of one another (i.e. $J_{AA}\ll0, J_{AB}\approx0$), although there is no globally preferred direction, we may observe $S>0$ due to the nematic order within each species. However, in general the direction of nematic order of species A and B will not be the same, hence we measure the difference in their direction as follows: define $\beta = \theta^{\textrm{nem}}_A-\theta^\textrm{nem}_B ~(\mathrm{mod~\pi})$, where $\theta_\Omega^\textrm{nem}=\langle \theta_\Omega^\textrm{nem}(t) \rangle_t$ is the time-averaged mean nematic direction of particles in the set $\Omega$. Nematic angular averages were computed by applying $\mathrm{atan2}$ to the vectorial averages of unit vectors $\hat{\mathbf{q}}_i=(\cos2\theta_i, \sin2\theta_i)$ to preserve periodicity. Then the difference in nematic direction is $\delta\theta_{S}=\min\{\beta, \pi-\beta\}$, which ensures that $\delta\theta_S \in [0, \pi/2)$.

\subsection{Demixing order parameter}

The species demixing order parameter for particle $i$ is defined as the ratio of same species particles to the total number of particles within a particular neighborhood of particle $i$:
\begin{equation}
	d_i = \frac{1}{|\mathcal{D}_i|}\sum_{j\in\mathcal{D}_i} \delta_{s(i) s(j)},
\end{equation}
where $\mathcal{D}_i=\{j:|\rbf_i-\rbf_j|\leq r_d\}$ are the particles within distance $r_d$ of particle $i$, and $\delta_{s(i)s(j)}$ is a Kronecker delta, which is equal to one if and only if the particles $i$ and $j$ are of the same species. We then average over particles, time (in steady-state) and realizations to get the average demixing order parameter
\begin{equation}
\bar{d} = \left \langle \frac{1}{N} \sum_{i=1}^N d_i -\frac{1}{2}\right \rangle.
\end{equation} 
Note that we subtract $1/2$ as is commonly done to ensure that $\bar{d}=0$ for a completely mixed system, and $\bar{d}>0$ if there is interspecies demixing.
\newline
We here are concerned with demixing that occurs on a lengthscale smaller than the interaction radius (the distance between stripes of different species tends to be around $0.6\RI$), thus we chose $r_d=\RI/2$ so that in the striped phases, $\mathcal{D}_i$ is almost only comprised of particles of the same species. We found this value of $r_d$ to be more attuned to slight changes in local species demixing compared to using the whole interaction kernel ($r_d=\RI$), particularly when there is system-wide nematic order.

\subsection{Periodicity}
\begin{figure}[h!]
	\includegraphics[width=0.6\linewidth]{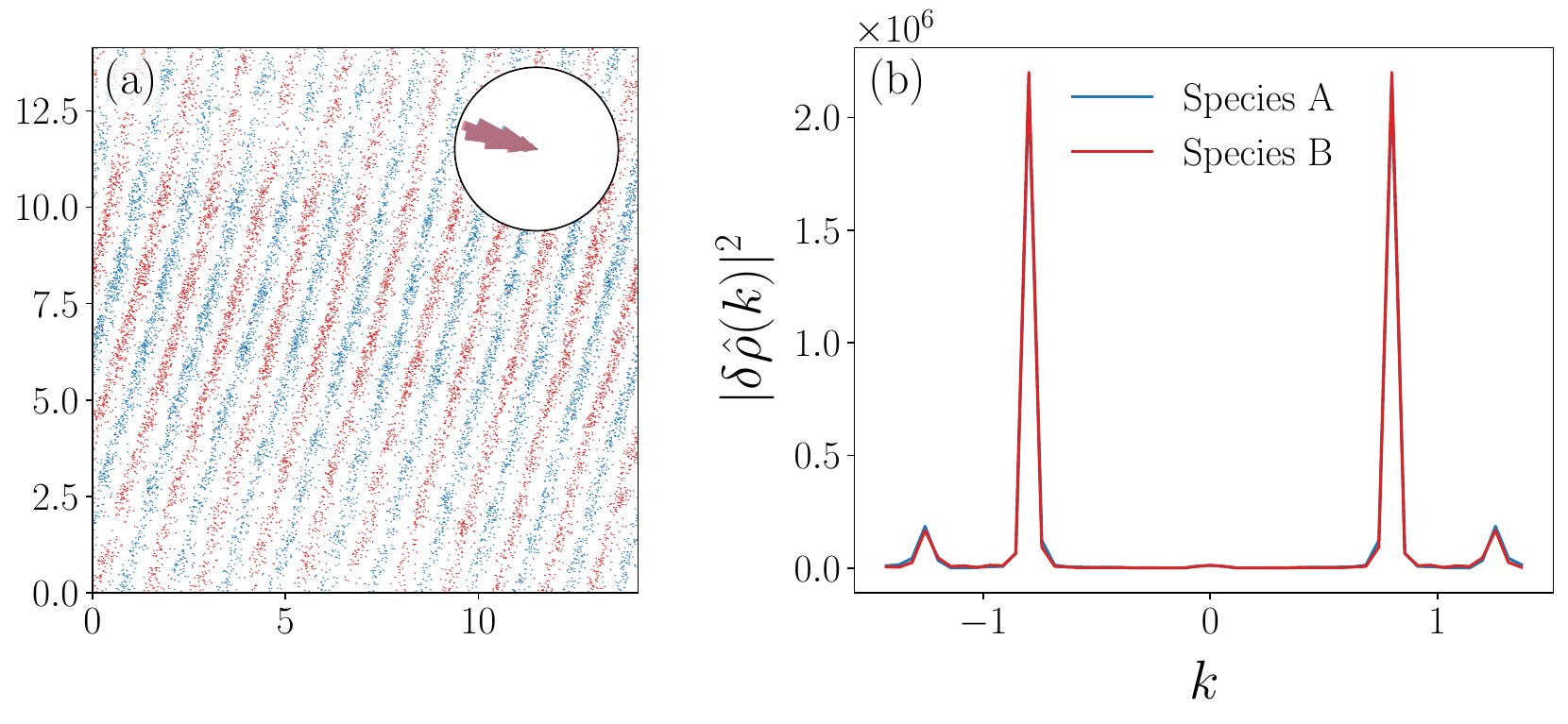}
	\caption{(a) Snapshot for a system with flocking stripes and (b) its power spectrum of the density fluctuations, projected onto direction of travel. Observe the overlapping power spectra for each species, with large peaks at $|k|\approx 0.8$, indicating that this system displays spatial periodicity at this wave number for both species. Simulation parameters are $N=2\times10^4$, $\rho=100$, $D_r=0.2$, $\gAA=-\gAB=20$.}
	\label{fig:power_spectrum}
\end{figure}

For each species, we project particles' position onto a line parallel to the mean direction of motion (or nematic direction), and measure the local density along this line $\rho(x_\parallel)$. In the banded periodic phases, the local density exhibits a periodic pattern along the direction of travel. To provide a robust measure of the presence (or absence) of periodicity, we look at the power spectrum of the density fluctuations $|\delta\hat{\rho}(k)|^2$, where $\delta\hat{\rho}$ is the Fourier transform of the density fluctuations $\delta\rho(x_\parallel)=\rho(x_\parallel) - \rho_0$. Peaks in this power spectrum denote the existence of periodic structures in the density fluctuations. As shown below, we expect in particular the largest peak to occur at a wave number $k\approx1/\lambda=1/(1.23 \sigma_I)$. In practice, we check for peaks above a chosen power threshold of $10^4$ in the interval $[0.5/\sigma_I,1/\sigma_I]$ in at least one of the species, in which case the system is classed as periodic (polar or nematic). An example of the power spectrum for a system with flocking stripes is shown in Fig.~\ref{fig:power_spectrum}.

\vspace{2cm}
%%%%%%%%%%%%%%%%%%%%%%%%%%%%%%%%%%%%%%%%%
\setlength{\tabcolsep}{10pt}
\begin{table}[h!]
\centering
\begin{tabular}{l|c|c|c|c|c|c|c|c}
& $\Psi$ & $\Psi_A,~\Psi_B$ & $|\delta \theta_{\Psi}|$ & $S$ & $S_A,~S_B$ & $|\delta \theta_{S}|$ & $\overline{d}$ & Periodicity \\ \hline \hline

Parallel flocking 
& $>0$ & $>0$ & $0$ & $>0$ & $>0$ & $0$ & $0$ & No \\ \hline

Antiparallel flocking 
& $0$ & $>0$ & $>0$ & $>0$ & $>0$ & $0$ & $0$ & No \\ \hline

Flocking stripes 
& $>0$ & $>0$ & $0$ & $>0$ & $>0$ & $0$ & $>0$ & Yes \\ \hline

Antiparallel flocking stripes 
& $0$ & $>0$ & $>0$ & $>0$ & $>0$ & $0$ & $>0$ & Yes \\ \hline

Nematic stripes (demixed) 
& $0$ & $0$ & -- & $>0$ & $>0$ & $0$ & $>0$ & Yes \\ \hline

Nematic stripes (mixed) 
& $0$ & $0$ & -- & $>0$ & $>0$ & $0$ & $0$ & Yes \\ \hline

Independent flocking 
& $0^*$ & $>0$ & $>0$ & $0$ & $0$ & -- & $0$ & No \\ \hline

Independent nematic ordering 
& $0$ & $0$ & -- & $0^*$ & $>0$ & $>0$ & $>0$ & Yes \\ \hline

Disordered 
& $0$ & $0$ & -- & $0$ & $0$ & -- & $0$ & No \\ \hline

\end{tabular}
\caption{Classification of emergent phases. *Note that the absence of global polar or nematic order does not necessarily mean that $\Psi=0$ or $S=0$, for all realizations; indeed, non-zero global order may emerge as a side effect of having species polar or nematic ordering in independent directions. We therefore check the difference in the preferred direction of travel of each species for these systems as these will be different in general.}
\label{tab:classification}
\end{table}
%%%%%%%%%%%%%%%%%%%%%%%%%%%%%%%%%%%%%%%%%

\clearpage
\section{Structure factor} \label{sec:structure_factor}
\begin{figure}[h!]
	\includegraphics[width=\linewidth]{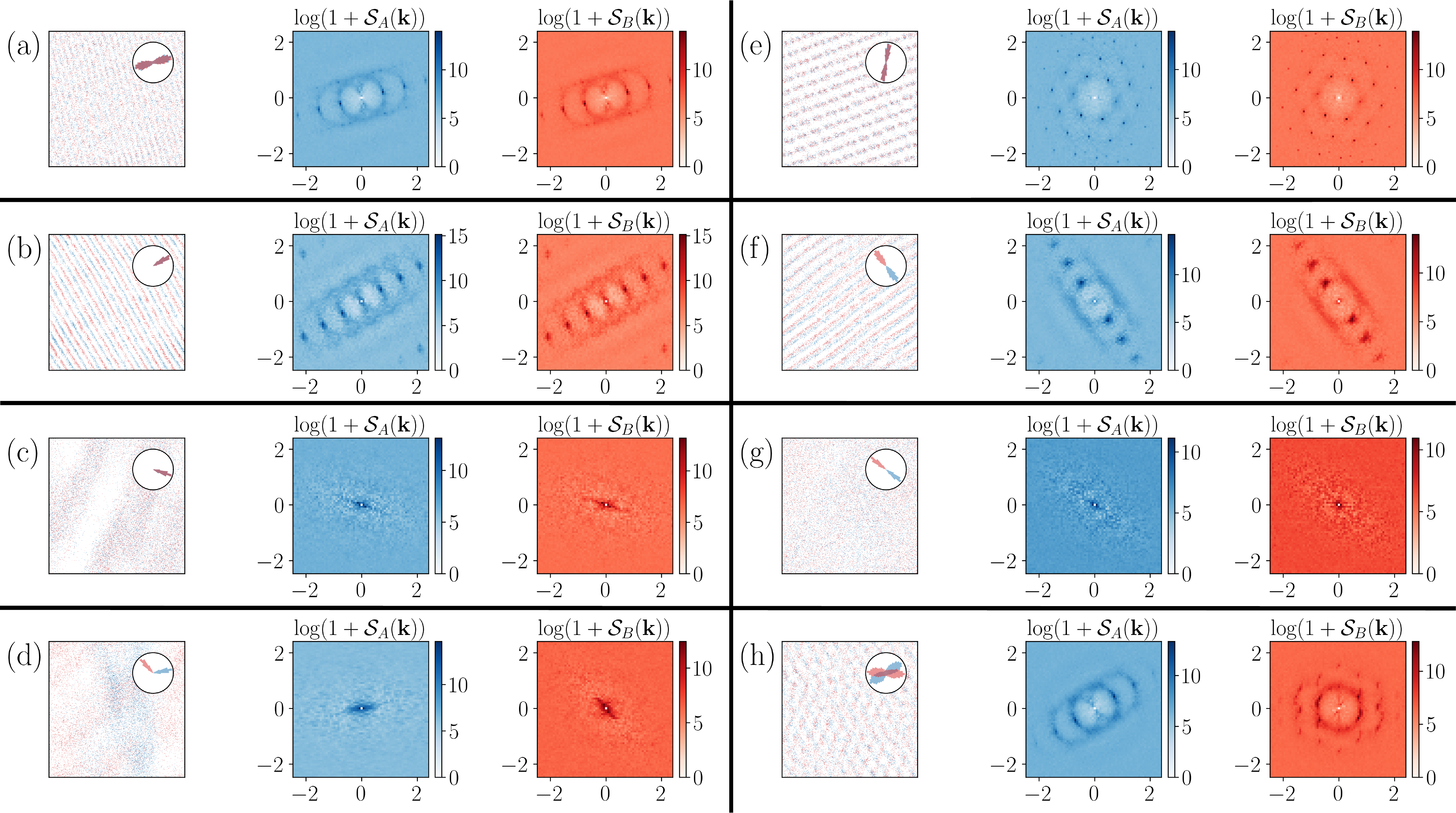}
	\caption{Structure factor for each phase, matching the phases and parameters in main text Fig.~1(a)--(h). Structure factors were calculated over $t=200$ to $220$ at increments of $dt=0.1$.}
	\label{fig:structure_factor}
\end{figure}

The structure factor is the Fourier transform of the two-point density correlation function; it is particularly useful when trying to understand the spatial structuring of a system and highlight any periodicity. First, we coarse-grain our system into a square grid with unit cell length $\ell=0.2$. We then compute the local densities in each of the grid cells, $\rho(\rbf,t)$, where $\rbf$ is the center of the grid cell. Using a fast Fourier transform, we compute the Fourier transform $\delta\hat{\rho}(\mathbf{k},t)$ of the density fluctuations $\delta\rho(\rbf, t)=\rho(\rbf,t) - \langle \rho(\rbf,t)\rangle_{\rbf}$. The structure factor is then defined as 
\begin{equation}
	\mathcal{S}(\mathbf{k}) = \langle |\delta\hat{\rho}(\mathbf{k}, t)|^2 \rangle_t,
\end{equation}
where $\langle \cdots \rangle_t$ is an average over time in steady-state. It can be calculated within each species separately $\mathcal{S}_A$, $\mathcal{S}_B$. In Fig.~\ref{fig:structure_factor}, we show the structure factors obtained for each of the observed phases. In the periodic systems, we observe a large peak in the direction of polar or nematic order at $|\mathbf{k}| \approx 0.8$. The wavelength of the flocking stripes (Fig.~\ref{fig:structure_factor}(a)) is discussed further in Section.~\ref{sec:wavelength}; however, it is useful to note that the same wavelength ($\lambda=1/k\approx 1.23$) appears in the other periodic phases as well [Fig.~\ref{fig:structure_factor}(b), (e), (f), (g)].

Here, note that we represent $\log (1+\mathcal{S}(\mathbf{k}))$ for ease of visualization as the peaks present in the structure factor can overpower the rest of the power spectrum.

\section{Hyperuniformity of the disordered state}
\begin{figure}
	\includegraphics[width=0.8\linewidth]{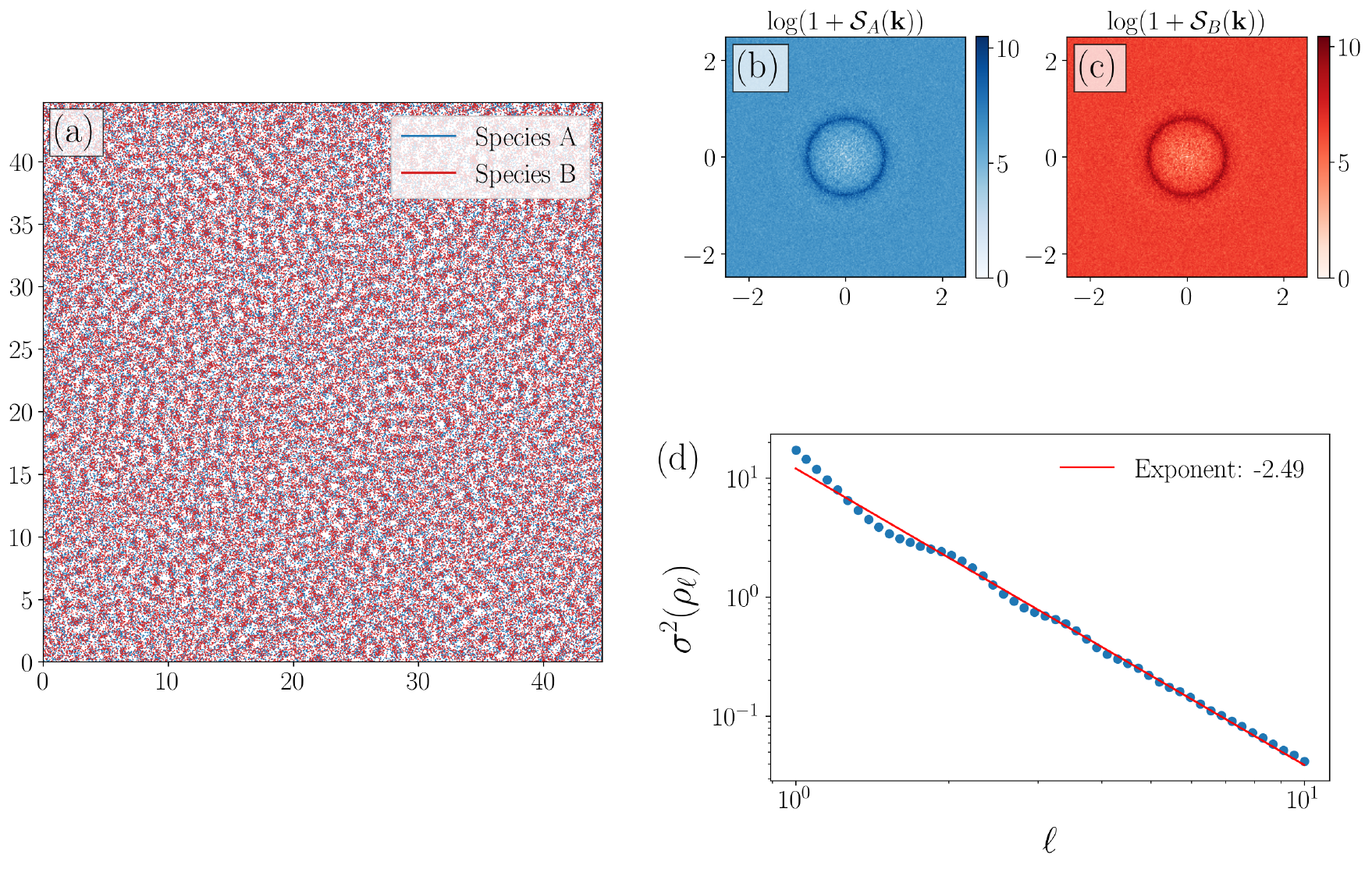}
	\caption{Hyperuniformity in the disordered phase. (a) Snapshot of a system with $N=2\times10^5$, $\rho=100$, $D_r=0.2$, $\gAA=-20$, $\gAB=-14$. (b)--(c) Structure factor for species A and B. (d) Variance of density $\rho_\ell$ as a function of the probing window size $\ell$. A power law is fitted, giving an exponent of $-2.49$.}
	\label{fig:hyperuniform}
\end{figure}

Although no directed motion was found in the disordered phase, the influence of highly anti-aligning intraspecies couplings can lead to spatial correlations. This becomes evident when simulating large systems [Fig.~\ref{fig:hyperuniform}(a)]. The structure factors for each species are plotted in Fig.~\ref{fig:hyperuniform}(b)--(c). For this system, we observe a sharp peaked ring (i.e. a peak at constant $|\mathbf{k}|$), implying that the system possesses one characteristic finite lengthscale.
Interestingly, the peak occurs at roughly $k=|\mathbf{k}|\approx0.8$, which leads to the same wavelength $\lambda=1/k$ measured for the flocking stripes in Section~\ref{sec:wavelength} (we work here in units of $\sigma_I =1$). Interestingly, this emergent length scale seems to appear in systems even without collective motion or demixing.

To quantify the spatial structuring further, we detect the presence of hyperuniformity \cite{torquato2018} in the system by calculating the variance of the local density $\sigma^2(\rho_\ell)$, where $\rho_\ell$ is the density of particles found in a randomly placed disk of radius $\ell$ (the probing window). The variance of the local density $\rho_\ell$ is expected to scale with the size of the probing window as
\begin{equation}
\sigma^2(\rho_\ell) \sim \ell^{-\lambda}.
\end{equation}
For Poisson point processes, the scaling exponent is expected to be $\lambda = d$ (here, $d=2$). If $\lambda>d$, then the density fluctuations are decaying faster than that of uniform systems, such a system is called hyperuniform. Finally, if $\lambda<d$, then one observes enhanced density fluctuations, the system is structured in space, and we can call our system clustered. For the parameters in question, we measured a scaling exponent of $\lambda=2.49>d$ [Fig.~\ref{fig:hyperuniform}]. We can therefore conclude that this disordered phase is hyperuniform, with a high degree of spatial structuring.

\clearpage
\section{Finite size analysis}
\begin{figure}[h!]
	\includegraphics[width=0.77\linewidth]{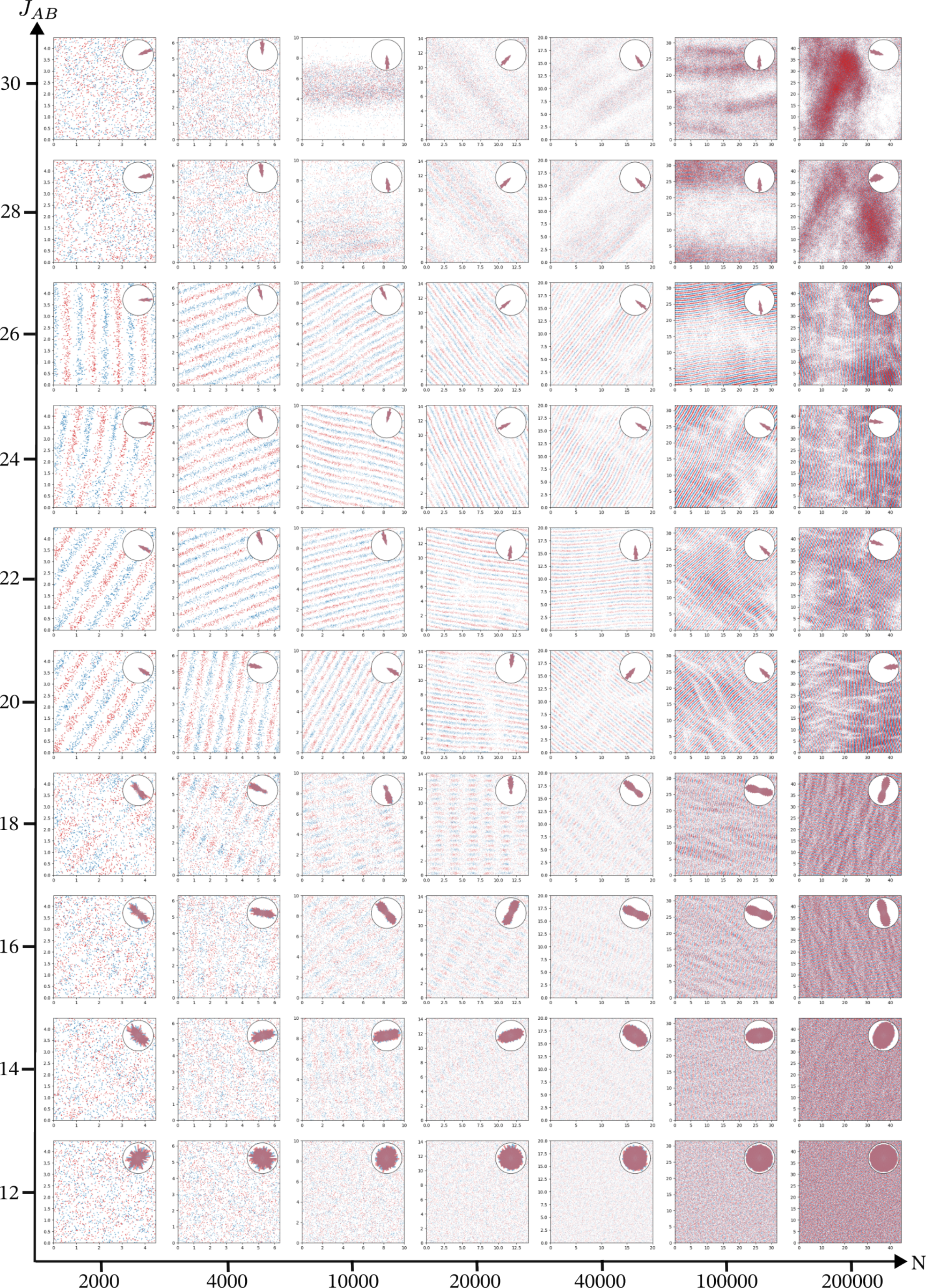}
	\caption{Snapshots at increasing system size with $\gAA=-20$ and $\gAB\in[12,30]$. All observed phases persist up to $N=2\times10^5$. Other simulation parameters are $\rho=100$ and $D_r=0.2$.}
	\label{fig:finite_size_pos}
\end{figure}

\begin{figure}
	\includegraphics[width=0.77\linewidth]{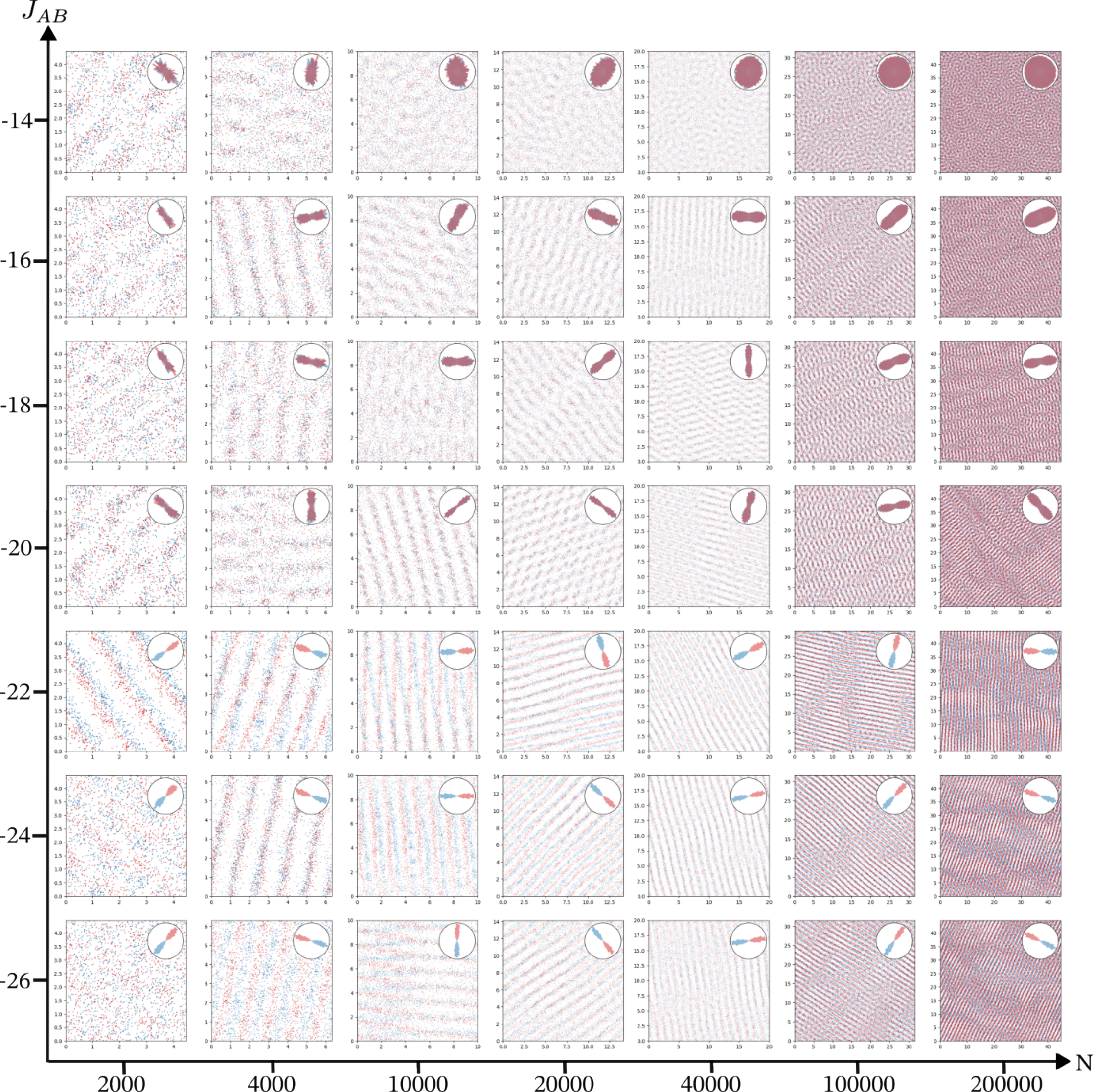}
	\caption{Snapshots at increasing system size with $\gAA=-20$ and $\gAB\in[-26,-14]$. All observed phases persist up to $N=2\times10^5$. Other simulation parameters are $\rho=100$ and $D_r=0.2$.}
	\label{fig:finite_size_neg}
\end{figure}

We performed simulations for various system sizes, $2000 \le N \le 2\times 10^5$, to ascertain the robustness of the observed phases. Simulation snapshots are reported in Fig.~\ref{fig:finite_size_pos} and in Fig.~\ref{fig:finite_size_neg} at fixed $\gAA=-20$, for different system sizes $N$ and varying $\gAB$. Although they took longer to reach a steady-state in the largest system sizes, we found all phases to persist even in the largest systems tested here. 

\clearpage

\section{Additive Sine Model}

In the main text, our model used a so-called mean-sine alignment, in which the torque applied on a given particle is divided by the number of particles within the interaction radius $n_i$. This is commonly done in Vicsek-like models to prevent excessive clumping due to the point-like nature of the particles. Here, we verify that the phases persist in an additive-sine alignment model, with Langevin equations:
\begin{subequations}
	\begin{align}
		\dot{\mathbf{r}}_i &= v_0 \hat{\mathbf{p}}_i,  \\
		\dot{\theta}_i &= \sum_{j\in\mathcal{N}_i} J_{s(i), s(j)} \sin(\theta_j-\theta_i) + \sqrt{2D_r} \ \xi_i.
	\end{align} \label{eq:langevin_AS}
\end{subequations}

\begin{figure}[h!]
	\includegraphics[width=0.8\linewidth]{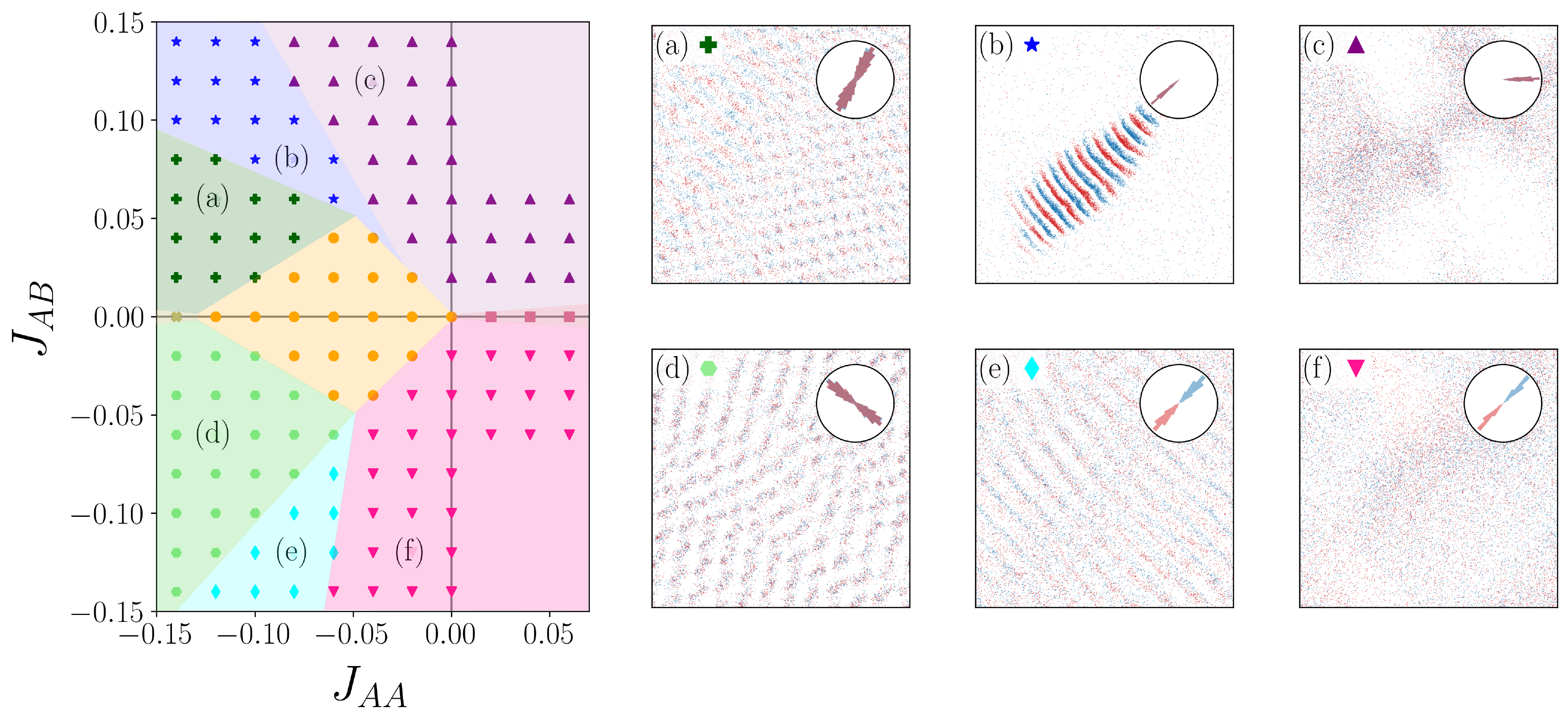}
	\caption{Additive sine phase diagram for $N=20000$ at $\rho=100$, $D_r=0.2$, and using timestep $\Delta t =5\times 10^{-3}$. Phases were classified numerically, with data averaged over 10 independent realizations for each data point. The phases are: Flocking stripes (dark-blue star); Nematic stripes, phase separated (light-green plus); Flocking (purple up-triangle); Flocking, independently (pink square); Nematic ordering, independently (khaki cross); Nematic stripes, mixed (dark-green hexagon); Antiparallel polar stripes (pink diamond); Antiparallel flocking (light-blue down-triangle); Disordered (orange circle). Phase lines were added manually. Example snapshots are: (a) $\gAA=-0.12$, $\gAB=0.06$; (b) $\gAA=-0.08$, $\gAB=0.08$;  (c) $\gAA=-0.04$, $\gAB=0.12$; (d) $\gAA=-0.12$, $\gAB=-0.06$; (e) $\gAA=-0.08$, $\gAB=-0.12$; (f) $\gAA=-0.02$, $\gAB=-0.12$. A polar histogram of the particle directions is shown in the upper right corner for each snapshot.}
	\label{fig:phase_diagram_AS}
\end{figure}

The phase diagram in the ($\gAA, \gAB$)-plane is shown in Fig.~\ref{fig:phase_diagram_AS}, along with snapshots of some of the phases.  The phase diagram remains remarkably qualitatively similar to that presented in the main text, albeit with a rescaling of $\gAA$ and $\gAB$ by $n_i$. This shows that the emergent phases we describe in the main text are robust to the exact definition of the alignment interactions. For these systems, the density is taken to be $\rho=100$; we argue that for a homogeneous system we expect any particle to interact with approximately 300 particles (although, they may interact with more neighbors in a clustered configuration). Therefore, the coupling values plotted are around 300 times smaller than those in the main text.
The only other main difference we observed is slightly more clustered states which is consistent with previous literature on the mean- and additive-sine models. For instance, the flocking stripes phase in Fig.~\ref{fig:phase_diagram_AS}(b) is now clustered rather than showing system-wide traveling bands, despite using the same density as in the main text.

\section{Nature of the flocking stripes}

\subsection{Wavelength} \label{sec:wavelength}
We compute the wavelength using the Fourier transform of the density fluctuations. First, for each species, we compute the structure factor (as in Section \ref{sec:structure_factor}), which we then project onto the mean direction of travel, giving a power spectrum in terms of $k_\parallel$. Here, we compute the dominant wavenumber (away from $k=0$) to calculate the species wavelength as $\lambda = 1/k_{max}$. We averaged over the species and different realizations with the same parameters to compute $\lambda$.
Some variability in $\lambda$ is expected for smaller system sizes with system spanning stripes, as the system always has to maintain an integer number of stripes, due to periodicity. 

\begin{figure}
	\includegraphics[width=0.8\linewidth]{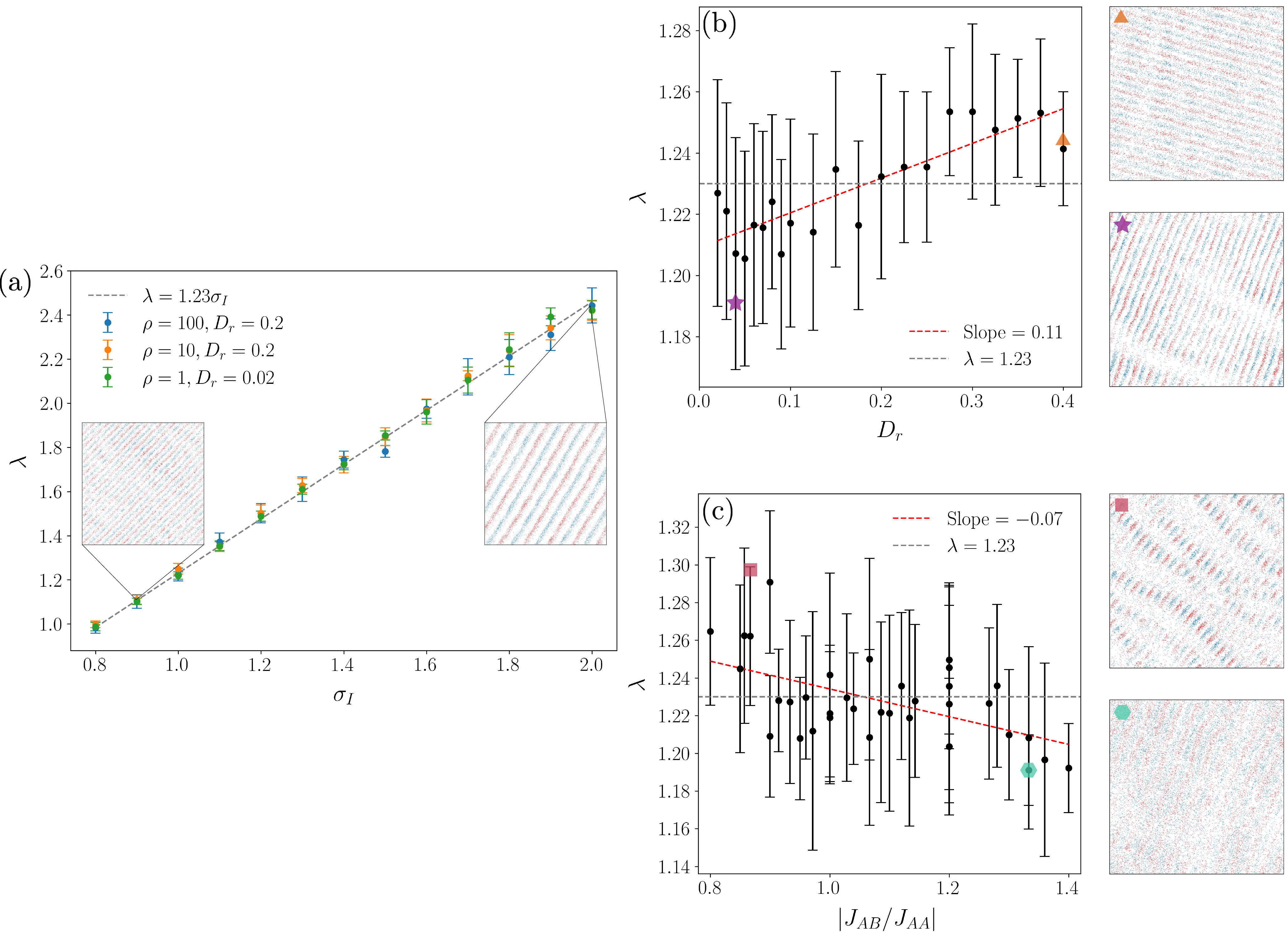}
	\caption{(a) Same species wavelength between bands $\lambda$ measured for different $\sigma_I$. Data are averaged over time and realizations, with error bars showing the variation between realizations. Data are plotted for $\rho=100$, $D_r=0.2$ (blue); $\rho=10$, $D_r=0.2$ (orange); $\rho=1$, $D_r=0.02$ (green). Note that noise had to be decreased for $\rho=1$ to enable flocking stripes to form.
	Wavelength is directly proportional to $\sigma_I$, with the $\lambda=1.23 \sigma_I$ line of best fit shown in gray. Example snapshots are shown for $\rho=100$ at $\sigma_I=0.9$ and $\sigma_I=2.0$. Other simulation parameters were $N=2\times 10^4$, $\gAA=-20$, and $\gAB=20$. 
	(b) Wavelength between bands of same species $\lambda$ as a function of noise parameter $D_r$. Data are averaged over time and realizations, with error bars showing the variation between realizations. There is a slight increase in wavelength as noise increases; a linear fit is shown in red with a slope of $0.11$. A gray line is drawn at $\lambda=1.23$ as a visual guide.
	Example snapshots are shown for $D_r=0.05$ (orange triangle) and $D_r=0.4$ (purple star). Other simulation parameters were $N=2\times 10^4$, $\rho=100$, $\gAA=-20$, $\gAB=20$, and $\sigma_I=1$. (c) Wavelength between bands of same species $\lambda$ as a function of the coupling ratio $|\gAB/\gAA|$. Data are averaged over time and realizations, with error bars showing the variation between realizations. There is a slight decrease in wavelength as the coupling ratio increases; a linear fit is shown in red with a slope of $-0.07$. A gray line is drawn at $\lambda=1.23$ as a visual guide.
	Example snapshots are shown for $(\gAA,\gAB)=(-30,40)$ (pink square), and $(\gAA,\gAB)=(-30,26)$ (turquoise hexagon). Other simulation parameters were $N=2\times 10^4$, $\rho=100$, $D_r=0.2$, and $\sigma_I=1$.}
	\label{fig:wavelength_Rp}
\end{figure}

In Fig.~\ref{fig:wavelength_Rp}(a), we plot the wavelength as a function of $\sigma_I$, at different densities.
At $\rho=100$ (the density used for the phase diagram in Fig.~1 of the main text), we find an average wavelength of $\lambda=(1.228 \pm 0.005) \sigma_I$.
At lower densities, the same linear dependence on $\sigma_I$ is found, with $\lambda\approx 1.23 \sigma_I$. We conclude that the wavelength of the stripes is independent of the system density. 

In terms of the other simulation parameters, we observe that the wavelength displays a weak dependence in the noise $D_r$ and coupling ratio $|\gAB/\gAA|$. At fixed coupling values and density, we find a small positive correlation of wavelength with the rotational diffusion coefficient $D_r$ [Fig.~\ref{fig:wavelength_Rp}(b)]. As noise decreases, the bands appear to become thinner and the local density inside the bands increases, which cause the stripe separation to decrease slightly.
At fixed noise and density, we find a small correlation of wavelength with the ratio of the intra- and interspecies coupling parameters [Fig.~\ref{fig:wavelength_Rp}(c)]. The coupling ratio is lower close to the boundary with the demixed nematic stripes phase and the particles tend to form tighter and narrower clusters, but with a larger stripe separation. Whereas for higher coupling ratios (close to the boundary with the homogeneous parallel flocking phase) we observe looser system-spanning stripes with a smaller stripe separation.
We note, however, that there is variation between realizations, and the mean value of $\lambda=1.23$ is within the error bounds for the majority of points in Fig.~\ref{fig:wavelength_Rp}(b)--(c).

\subsection{Velocity correlations}

\begin{figure}
	\includegraphics[width=0.5\linewidth]{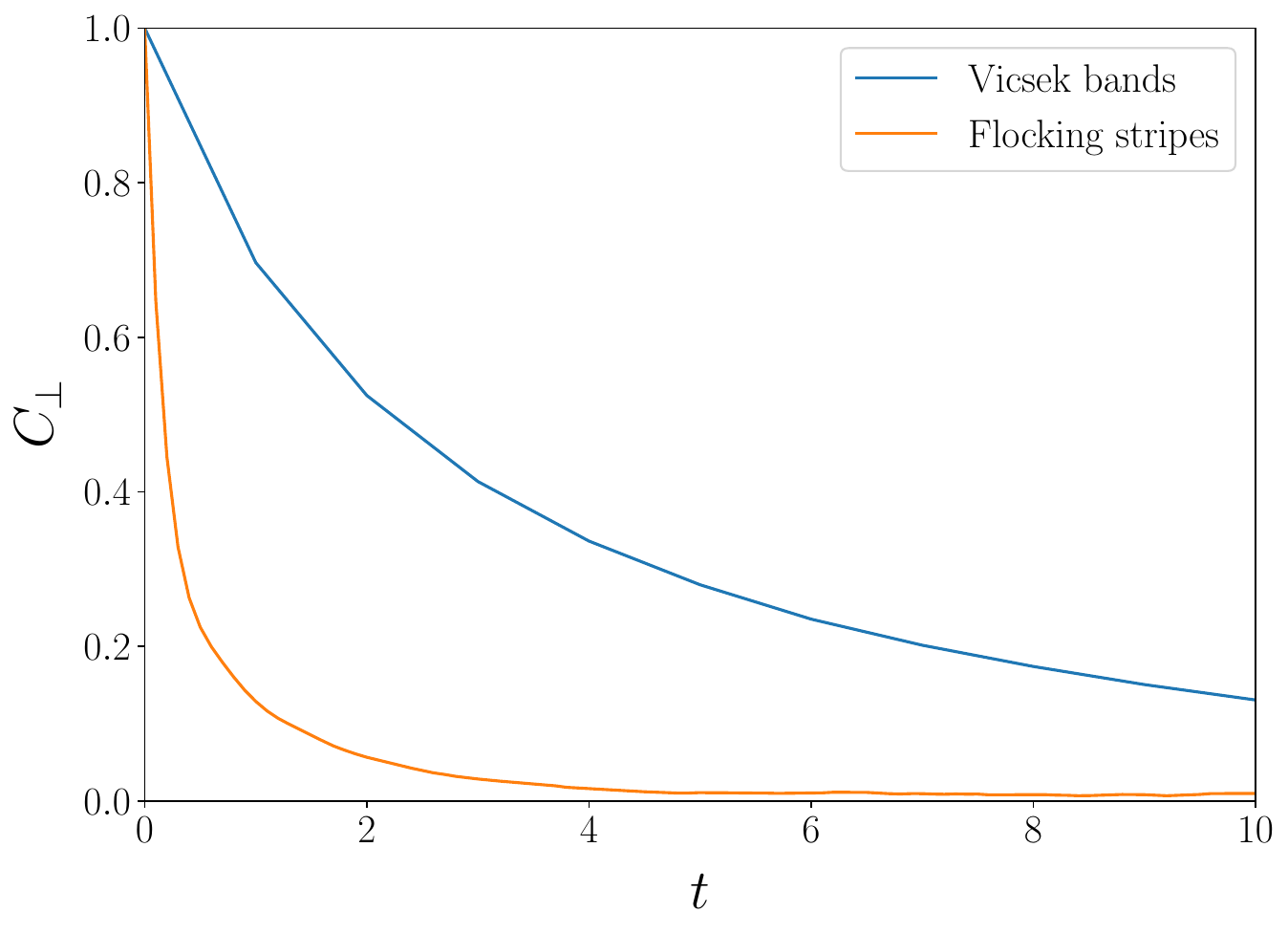}
	\caption{Velocity correlations perpendicular to the direction of travel for a system in either the coexistence band regime of the Vicsek model, or the flocking stripe regime of the two species model. Simulation parameters for one species Vicsek bands: $N=10^4$, $\rho=1$, $\sqrt{2D_r}=0.65$, $J=1$, in a slab domain $L_x/L_y=5$. Simulation parameters for two species flocking stripes: $N=10^4$, $\rho=200$, $D_r=0.2$, $\gAA=-\gAB=-30$ in a square domain.}
	\label{fig:vel_corr_t}
\end{figure}

We measure the two-time correlator for the velocity perpendicular to the mean direction of polar order
\begin{equation}
	C_\perp(t) = \frac{\langle\vbf_\perp^{(i)}(\tau)\cdot \vbf_\perp^{(i)}(\tau+t) \rangle_{i,\tau}}{\langle |\vbf_\perp^{(i)}(\tau)|^2 \rangle_{i,\tau}},
\end{equation}
which has been normalized such that $C_\perp(0)=1$. We choose a system in the flocking stripes phase, as well as a system in the coexistence regime of the Vicsek model, displaying traveling bands. Their velocity correlations $C_\perp(t)$ have been plotted in Fig.~\ref{fig:vel_corr_t}. The correlations decay extremely quickly in the flocking striped system, especially compared to the Vicsek bands. This tells us that, once they are formed, the system-spanning flocking stripes are extremely stable, with little deviation from the mean direction of travel. 

\section{Non-symmetric parameters}
All results reported in the main text are for the fully symmetric case in which $N_A=N_B=N/2$ and $J_{AA}=J_{BB}$. Here, we wish to test the robustness of the flocking stripes to species heterogeneity.

\begin{figure}
\includegraphics[width=\linewidth]{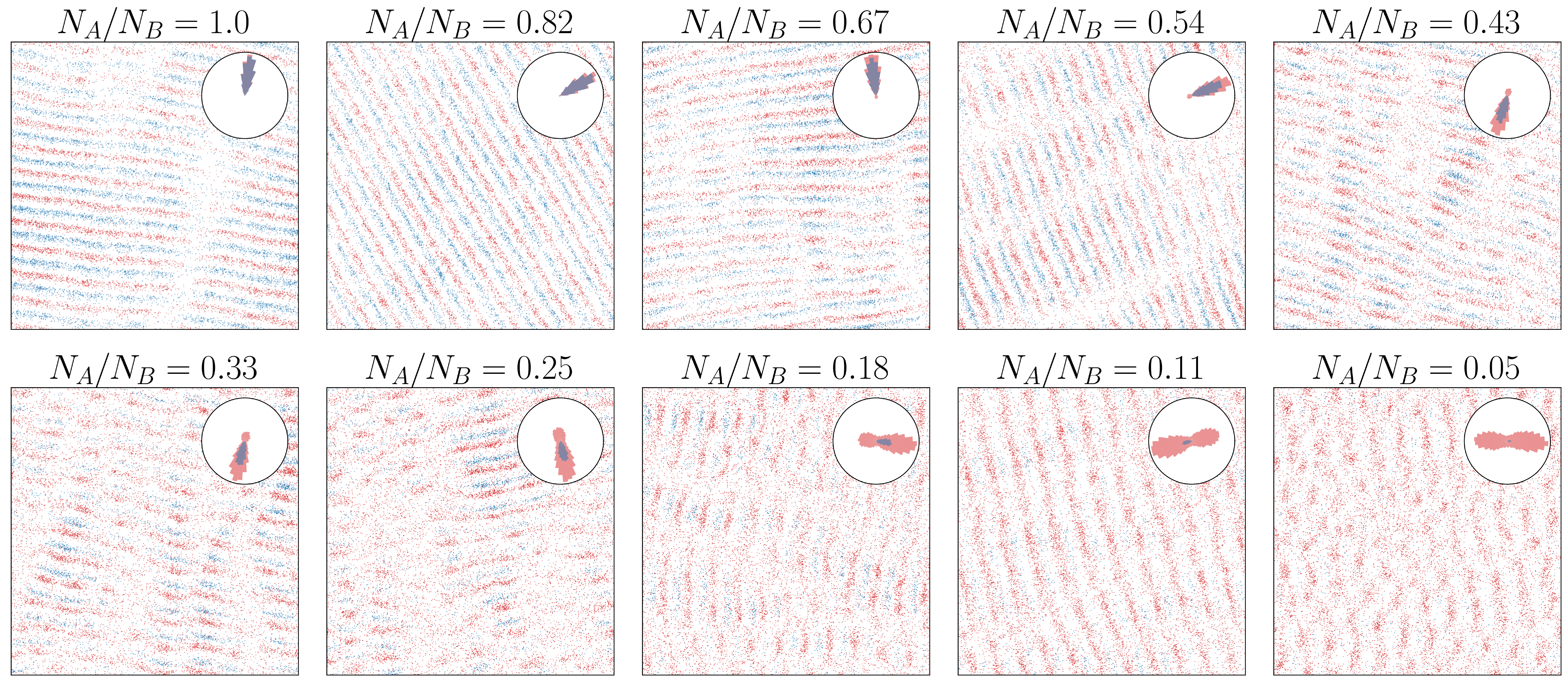}
\caption{Snapshots of simulations, varying the species ratio. We kept the total particle number constant $N=N_A+N_B=2\times10^4$. Other simulation parameters were $\rho=100$, $D_r=0.2$, $J_{AA}=J_{BB}=-20$, and $J_{AB}=J_{BA}=20$.}
\label{fig:vary_NA}
\end{figure}

Firstly, we varied the particle species ratio $N_A:N_B$, while keeping the total number of particles $N$ and system density constant. Snapshots are shown in Fig.~\ref{fig:vary_NA}. Starting from the symmetric case of $N_A/N_B=1$, we have chosen parameters that display system-spanning flocking stripes. As the relative proportion of species $A$ particles is decreased, parallel flocking stripes persist for a remarkably large range of ratios. 
As the fraction decreases below $N_A/N_B=0.5$ (i.e. a ratio of 1:2), some excess species B particles (red) begin to flock in the opposite direction, creating some nematic order in the system. Meanwhile, species $A$ (blue) maintain an alternating stripe pattern with species $B$ particles in at least part of the system box. Eventually the number of species $A$ is too small in this finite system to even support stripe formation; however, species $B$ particles have transitioned to an independent nematic striped / laned phase, as seen in the phase diagram when intraspecies coupling has a high degree of anti-alignment $J_{AA}\ll-1$.
Overall, flocking stripes persist for a large range of species proportions, even as asymmetric as a ratio of 1:4.

\begin{figure}
\includegraphics[width=\linewidth]{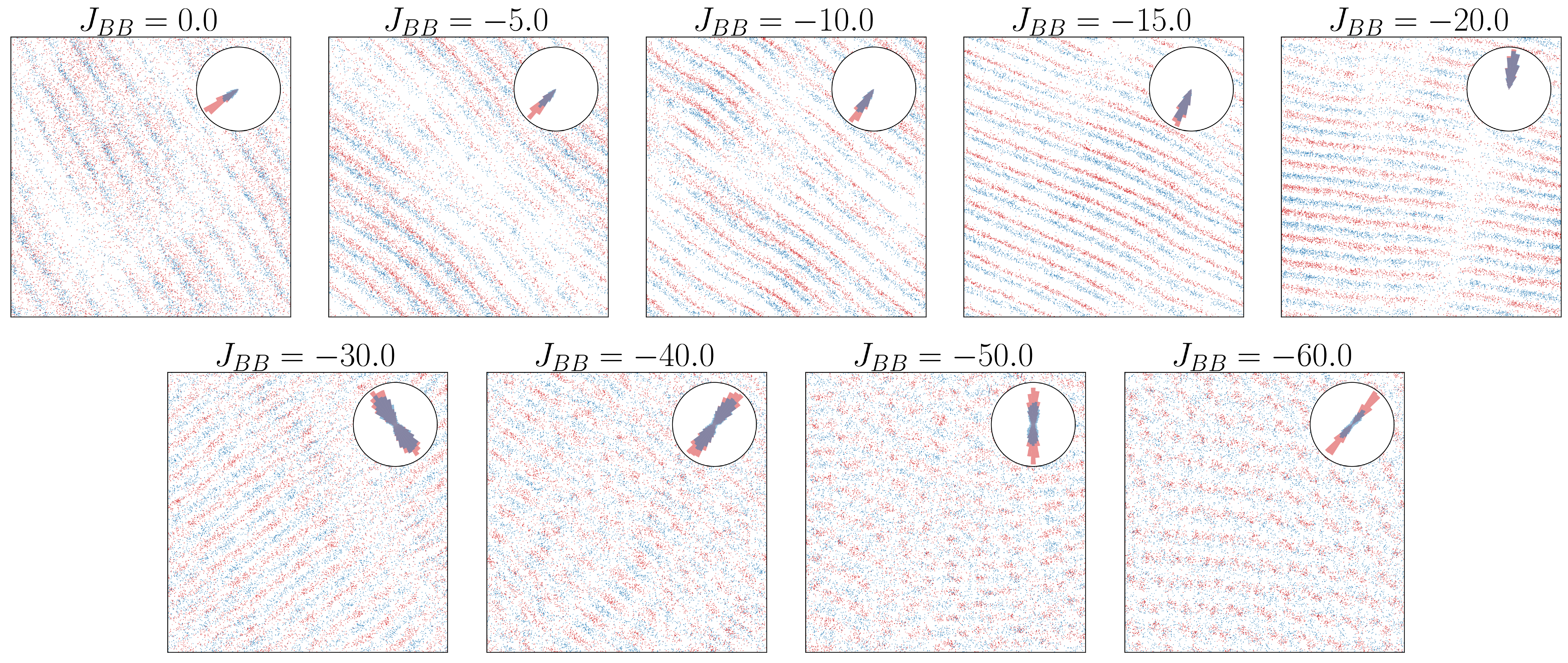}
\caption{Snapshots of simulations, varying intraspecies coupling for species B ($J_{BB}$). We kept $J_{AA}=-20$, $J_{AB}=J_{BA}=20$ constant.
Other simulation parameters were $N=2\times10^4$, $\rho=100$, and $D_r=0.2$.}
\label{fig:vary_JAA}
\end{figure}

Next, we will keep the species numbers equal but instead vary the self alignment couplings $J_{AA}$ and $J_{BB}$. In Fig.~\ref{fig:vary_JAA}, the values of $J_{AA}=-20$ and $J_{AB}=J_{BA}=20$ are kept constant, while we vary $J_{BB}$. The perfectly symmetric case is shown at $J_{BB}=-20$ with parallel flocking stripes.
As the magnitude of $J_{BB}$ is decreased, parallel flocking stripes persist, though appear to become less regular, with the red stripes (species $B$) closer to the blue stripe (species $A$) in front of it compared to behind. Species A flocking stripes even persist when $J_{BB}=0$, although species B no longer forms stripes due to the lack of intraspecies interaction. Species $B$ instead form a flocking liquid phase, supporting the microphase separation of species $A$ into flocking stripes.
If we instead look at increasing the magnitude of $J_{BB}$ such that its magnitude is larger than $|J_{AA}|=|J_{AB}|=|J_{BA}|=20$, we find that the strong intraspecies anti-alignment of species B dominates, with a transition to a demixed nematic stripes / lanes phase. 
Overall, we find that the flocking stripes are found even for a fairly large degree of intraspecies coupling asymmetry when they are less or equal in magnitude to the interspecies alignment.
Flocking stripes are lost when intraspecies anti-alignment of a single species dominates in magnitude over the other species and any interspecies alignment, instead displayed a nematic striped or laned phase.

\section{Other interaction kernels}

\begin{figure}
\includegraphics[width=\linewidth]{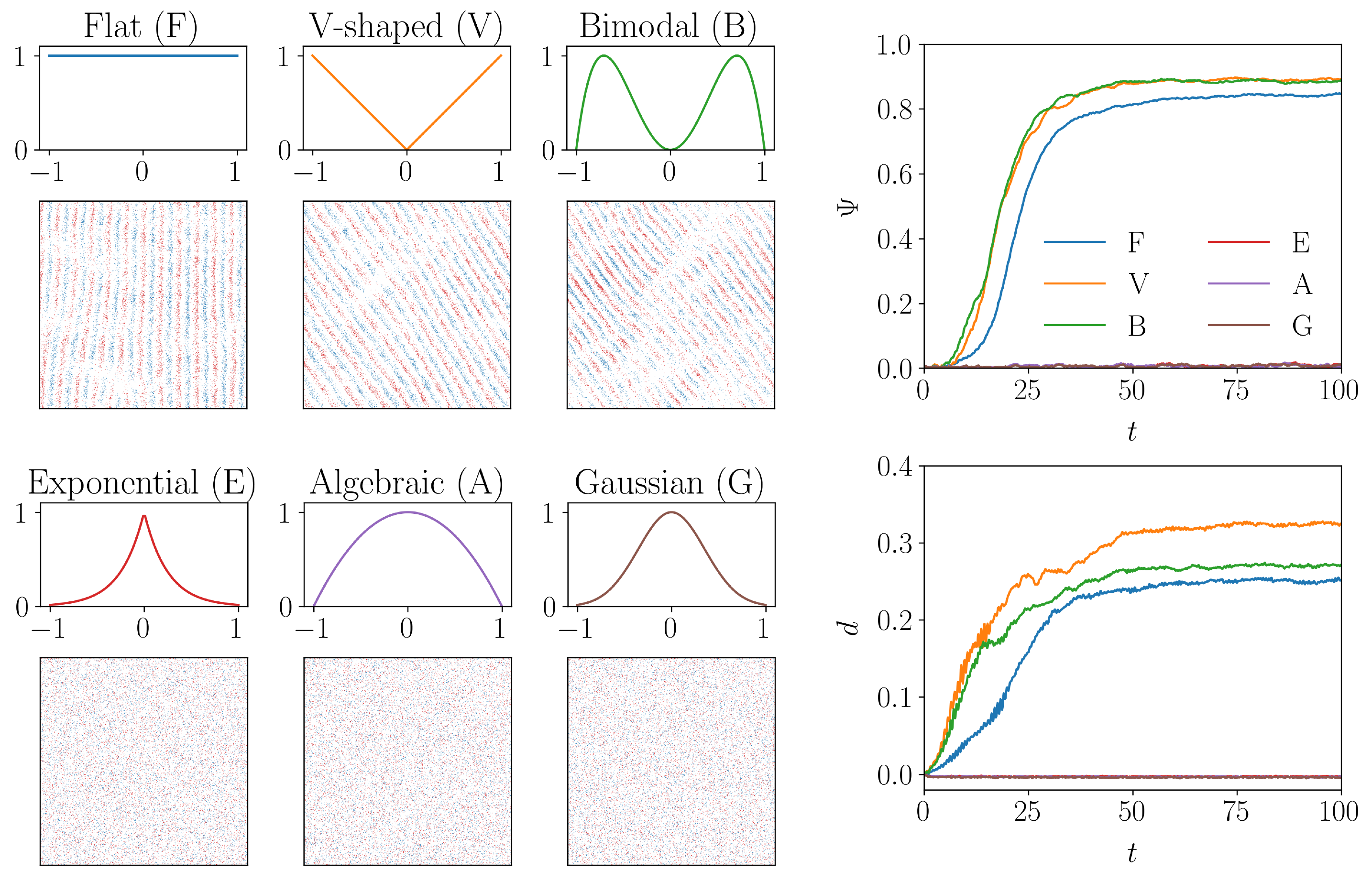}
\caption{Left: Various interaction kernels alongside snapshots from simulations using that kernel. The top three (Flat, V-shaped and Bimodal) show flocking stripes patterns, while the bottom three (Exponential, Algebraic and Gaussian) do not and are homogeneous disordered. Top right: Polar order parameter plotted against time for all six interaction kernels. Bottom right: Demixing order parameter plotted against time for all six interaction kernels. Only the Flat, V-shaped and Bimodal kernels display any polar order or demixing.}
\label{fig:vary_kernel}
\end{figure}

The mean field stability argument provided in the main text relies upon a metric interaction with radius $\RI$, at which there is a sharp discontinuity in the interaction. We wish to change this interaction to understand the stripe mechanism further, and to test its robustness.
We modify the orientation update to include an interaction kernel $K(r_{ij})$ that can be varied:
\begin{equation}
	\dot{\theta}_i = \frac{1}{n_i} \sum_{j\in\mathcal{N}_i} K(|\rbf_i-\rbf_j|) J_{s(i), s(j)} \sin(\theta_j-\theta_i) + \sqrt{2D_r} \ \xi_i,
\end{equation}
where the neighborhood $\mathcal{N}_i$ remains the same as in the main text and assume $\RI=1$. We tested the following kernels:
\begin{itemize}
	\item Flat: $K(x)=1$
	\item V-shaped: $K(x)=1-|x|$
	\item Bimodal (quartic): $K(x)=4x^2 - 4x^4$
	\item Exponential: $K(x)=e^{-4|x|}$
	\item Algebraic (quadratic): $K(x)=1-x^2$
	\item Gaussian: $K(x)=e^{-4x^2}$
\end{itemize}
Snapshots and order parameters from simulations of each kernel are shown in Fig.~\ref{fig:vary_kernel}. The `Flat' kernel case is the one used throughout the main text, with a sharp cut-off at $r_{ij}=\RI$. 

We experimented with some smooth kernels e.g. `Exponential' and `Gaussian' kernels, who peak at $r_{ij}=0$ and decay smoothly to almost zero at $r_{ij}=\RI$. However, these both seem to impede stripe formation, and remain in a disordered phase. Similarly, the `Algebraic' quadratic kernel, which is again unimodal around $r_{ij}=0$ and decays to zero, remains in a disordered state.
This is not completely unexpected, as the stripe pattern relies upon a balance of some same-species antiferromagnetic interactions close to $r_{ij}=0$, and slightly more opposite-species ferromagnetic interactions closer to the edges of the radius of interaction. However, if the weighting of particles around $r_{ij}=0$ is much larger than close to $r_{ij}=\RI$, then such a balance of species cannot be achieved.

If we instead used an interaction kernel with a larger weighting around the stripe separation of $\lambda/2\approx 0.6\RI$, then we expect the formation of stripes may actually be aided.
Testing both a `V-shaped' and quartic `Bimodal' kernel, this is indeed what we found. System-spanning stripes are recovered despite the non-constant interaction kernel. In fact, these kernels appear to speed up the formation of stripes, since the polar order and demixing parameters tend to start growing earlier in these simulations.
Furthermore, the bimodal kernel is an example of a kernel described by a continuous function, with the interaction decaying to zero at $r_{ij}=\RI$. 

Overall, these findings support our mean-field explanation that the formation and stability of the stripes relies upon there being more cross-species aligning interactions at a nonzero finite distance parallel to the direction of motion, compared to the weighting of same-species anti-aligning interactions around $r_{ij}\approx 0$.
The formation of stripes can even be sped up through interaction kernels that favor a particle separation of $\lambda/2$.

\clearpage
\section{Stripes with repulsion}
All simulations thus far have considered point-like particles. We finally consider systems in which particles interact via a purely soft repulsive force

\begin{subequations}
	\begin{align}
		\dot{\mathbf{r}}_i &= v_0 \hat{\mathbf{p}}_i + \sum_{j\in\mathcal{N}_i} \mathbf{F}_r(|\rbf_{ij}|),  \\
		\dot{\theta}_i &= \sum_{j\in\mathcal{N}_i} J_{s(i), s(j)} \sin(\theta_j-\theta_i) + \sqrt{2D_r} \ \xi_i,
	\end{align} \label{eq:langevin_rep}%
\end{subequations}%
where $\mathbf{F}_r(r)= - \nabla V(r)$ for a given potential $V(r)$. We argue that the formation of striped traveling bands requires a sufficiently soft potential; here, we choose a harmonic potential taking the form $V(r) = \kappa(\RR - r)^2$ for $r<\RR$ and $0$ otherwise, with stiffness parameter $\kappa$ and radius of repulsion $\RR$ acting as an effective particle diameter. The additive sine model for the alignment was used, as the repulsion already restricts the number of interaction neighbors, thereby preventing strong clumping.

\begin{figure}[h!]
	\includegraphics[width=0.4\linewidth]{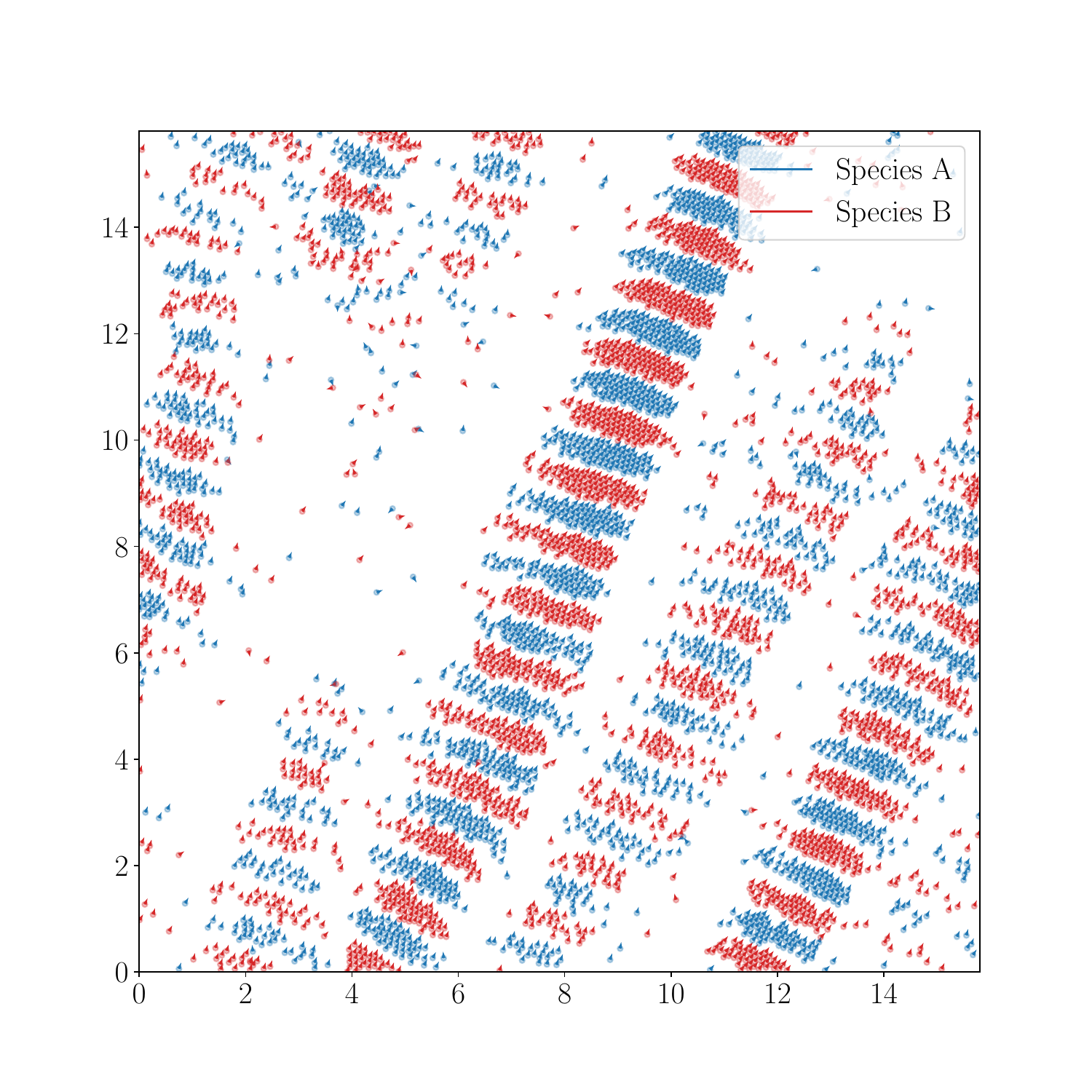}
	\caption{Snapshot of simulation with soft repulsion. Parameters: $N=5\times10^3$, $\rho=20$, $v_0=1$, $D_r=0.02$, $J_{AA}=1$, $J_{AB}=-1$, $\RI=1$, $\RR=0.1$, $\kappa=10$, $\Delta t=10^{-4}$.}
	\label{fig:repulsion_stripes}
\end{figure}

We find that the flocking stripes phase can be found with a soft repulsion. This requires $\sigma_R\ll \RI$, such that particles are still able to interact with a sufficient number of neighbors. An example snapshot is shown in Fig.~\ref{fig:repulsion_stripes}. We note, however, that these clustered configurations seem to be far more unstable, with even small fluctuations in density resulting in transient oscillatory patterns. 

